\documentclass[sigconf]{acmart} 

\usepackage{booktabs} 

\usepackage{microtype}
\usepackage{graphicx}
\usepackage{subfigure}
\usepackage{booktabs} 
\usepackage{makecell}
\usepackage{microtype}      
\usepackage{listings}
\usepackage{url} 
\usepackage{diagbox}

\usepackage{pdflscape}
\usepackage{mathrsfs}
\usepackage{paralist}
\usepackage{multirow}
\usepackage{mathtools,xparse}
\usepackage{enumitem}
\usepackage{amsmath}
\usepackage{bbold}
\usepackage{mathtools}
\usepackage{xcolor}
\usepackage{wrapfig}
\usepackage[normalem]{ulem}
\usepackage{amsthm}

\newcommand{\bit}{\begin{compactitem}}
\newcommand{\eit}{\end{compactitem}}
\newcommand{\ben}{\begin{compactenum}}
\newcommand{\een}{\end{compactenum}}
\newcommand{\mytag}[1]{{\bf#1}}

\newcommand{\model}{{NDCN\ }}
\DeclareMathOperator*{\st}{subject \ to \ }
\DeclarePairedDelimiter{\abs}{\lvert}{\rvert}
\DeclareMathOperator*{\softmax}{softmax}
\DeclareMathOperator*{\relu}{ReLU}

\newcommand{\hide}[1]{}

\DeclareMathOperator*{\argmin}{arg\,min}

\usepackage[ruled,vlined]{algorithm2e}

\theoremstyle{definition}

\theoremstyle{remark}

\setcopyright{acmlicensed}

\copyrightyear{2020}
\acmYear{2020}
\setcopyright{acmlicensed}\acmConference[KDD '20]{Proceedings of the 26th ACM SIGKDD Conference on Knowledge Discovery and Data Mining}{August 23--27, 2020}{Virtual Event, CA, USA}
\acmBooktitle{Proceedings of the 26th ACM SIGKDD Conference on Knowledge Discovery and Data Mining (KDD '20), August 23--27, 2020, Virtual Event, CA, USA}
\acmPrice{15.00}
\acmDOI{10.1145/3394486.3403132}
\acmISBN{978-1-4503-7998-4/20/08}

\hide{
\copyrightyear{2019} 
\acmYear{2019} 
\setcopyright{acmcopyright}
\acmConference[KDD '19]{The 25th ACM SIGKDD Conference on Knowledge Discovery and Data Mining}{August 4--8, 2019}{Anchorage, AK, USA}
\acmBooktitle{The 25th ACM SIGKDD Conference on Knowledge Discovery and Data Mining (KDD '19), August 4--8, 2019, Anchorage, AK, USA}
\acmPrice{15.00}
\acmDOI{10.1145/3292500.3330842}
\acmISBN{978-1-4503-6201-6/19/08}
}

\begin{document}
\title{
Neural Dynamics on Complex Networks}
 
\author{Chengxi Zang}
\affiliation{%
  \institution{Department of Population Health Sciences, Weill Cornell Medicine
}
}
\email{chz4001@med.cornell.edu}

\author{Fei Wang}
\affiliation{%
  \institution{Department of Population Health Sciences, Weill Cornell Medicine 
}
}
\email{few2001@med.cornell.edu}



\begin{abstract}
Learning continuous-time dynamics on complex networks is crucial for understanding, predicting and controlling complex systems in science and engineering. However, this task is very challenging due to the combinatorial complexities in the structures of  high dimensional systems, their elusive continuous-time nonlinear dynamics, and their structural-dynamic dependencies.
To address these challenges, we propose to combine Ordinary Differential Equation Systems (ODEs) and Graph Neural Networks (GNNs) to learn continuous-time dynamics on complex networks in a data-driven manner. We model differential equation systems by GNNs. Instead of mapping through a discrete number of neural layers in the forward process, we integrate GNN layers over continuous time numerically, leading to capturing continuous-time dynamics on graphs. Our model can be interpreted as a Continuous-time GNN model or a Graph Neural ODEs model.  
Our model can be utilized for  continuous-time network dynamics prediction, structured sequence prediction (a regularly-sampled case), and node semi-supervised classification tasks  (a one-snapshot case) in a unified framework.
We validate our model by extensive experiments in the above three scenarios. The promising experimental results demonstrate our model's capability of jointly capturing the structure and dynamics of complex systems in a unified framework.



\end{abstract}



\maketitle

\section{Introduction}
\label{sec:intro}

Real-world complex systems, such as brain \cite{gerstner2014neuronal}, ecological systems \cite{gao2016universal}, gene regulation \cite{alon2006introduction}, 
human health \cite{bashan2016universality}, and social networks \cite{li2019fates,lu2018collective,zang2016beyond,zang2018power,zang2019uncovering,zang2019dynamical}, etc., are usually modeled as complex networks and their evolution are governed by some underlying nonlinear dynamics \cite{newman2011structure}. Revealing these dynamics on complex networks modeled by differential equation systems is crucial for understanding the complex systems in nature. Effective analytical tools developed for this goal can further help us predict and control these complex systems. 
Although the theory of (nonlinear) dynamical systems has been widely studied in different fields including applied math \cite{strogatz2018nonlinear},
statistical physics \cite{newman2011structure},
engineering \cite{slotine1991applied}, ecology \cite{gao2016universal} and biology \cite{bashan2016universality}, these developed models are typically based on a clear knowledge of the network evolution mechanism which are thus usually referred to as {\em mechanistic models}. Given the complexity of the real world, there is still a large number of complex networks whose underlying dynamics are unknown yet (e.g., they can be too complex to be modeled by explicit mathematical functions). At the same time, massive data are usually generated during the evolution of these networks. Therefore, modern data-driven approaches are promising and highly demanding in learning dynamics on complex networks. 

The development of a successful data-driven approach for modeling continuous-time dynamics on complex networks is very challenging: the interaction structures of the nodes in real-world networks are complex and the number of nodes and edges is large, which is referred to as the high-dimensionality of  complex systems; the rules governing the dynamic change of nodes' states in complex networks are continuous-time and nonlinear; the structural and dynamic dependencies within the system are difficult to model by explicit mathematical functions. 
Recently, there has been an emerging trend in the data-driven discovery of
ordinary differential equations (ODE) or partial differential equations (PDE) to capture the continuous-time dynamics,
including sparse regression method \cite{mangan2016inferring,rudy2017data}, residual network \cite{qin2018data}, feedforward neural network
\cite{raissi2018multistep}, etc.
However, these methods can only handle very small ODE  or PDE systems which consist of only a few interaction terms. For example, the sparse-regression-based method \cite{mangan2016inferring} shows that its combinatorial complexity grows with the number of agents when building candidate interaction library.  
Effective learning of continuous-time dynamics on large networks which consist of tens of thousands of agents and interactions
is still largely unknown.


In this paper, we propose to 
combine Ordinary Differential Equation Systems (ODEs) and  Graph Neural Networks (GNNs) \cite{wu2019comprehensive} to learn non-linear, high-dimensional and continuous-time dynamics on graphs \footnote{We use graph to avoid the ambiguity with neural network. Otherwise, we use graph and network interchangeably to refer to linked objects.}. We model differential equation systems by GNNs to capture the instantaneous rate of change of  nodes' states.  
Instead of mapping through a discrete number of layers in the forward process of conventional neural network models \cite{lecun2015deep}, 
we \emph{integrate}  \cite{chen2018neural} GNN layers over continuous time rather than discrete depths, leading to a novel \emph{Continuous-time GNN} model.
In a  dynamical system view, the continuous depth can be interpreted as continuous physical time, and the outputs of any hidden GNN layer at time $t$ are instantaneous network dynamics at that moment. Thus we can also interpret our model as a \emph{Graph Neural ODEs} 
in analogy to the Neural ODEs \cite{chen2018neural}.
 \hide{
 In the backward learning process, we back-propagate the gradients of the supervised information w.r.t. the learnable parameters against the forward integration,
 leading to learning the differential equation system in an end-to-end manner. }
Besides, we further enhance our algorithm by learning the dynamics in a hidden space learned from the original space of nodes' states. We name our model Neural Dynamics on Complex Networks (\mytag{NDCN}). 

Our \model model can be used for following three tasks  in \emph{a unified framework}:
1) { (Continuous-time network dynamics prediction): Can we predict the continuous-time dynamics on complex networks at an arbitrary time}? 2) { (Structured sequence prediction \cite{seo2018structured}): Can we predict next few steps of structured sequences}?  3) { (Graph semi-supervised classification \cite{YangCS16,KipfW17}): Can we infer the  labels of each node given features for each node and some labels at one snapshot}?  The experimental results show that our \model can successfully finish above three tasks.
As for the task 1, our \model is first of its kind which learns continuous-time dynamics on large complex networks modeled by differential equations systems. As for the task 2, our \model achieves lower error in predicting future steps of the structured sequence with much fewer parameters than the temporal graph neural network models \cite{yu2017spatio,kazemi2019relational,narayan2018learning,seo2018structured}. As for the task 3, our \model learns a continuous-time dynamics to spread features and labels of nodes to predict unknown labels of nodes, showing very competitive performance compared to many graph neural networks \cite{KipfW17,thekumparampil2018attention,velivckovic2017graph}.
Our framework potentially serves as a unified framework to jointly capture the structure and dynamics  of complex systems in a data-driven manner.


It's worthwhile to summarize our contributions as follows:
\bit
    \item \mytag{A novel model:} We propose a novel model NDCN, which combines Ordinary Differential Equation Systems (ODEs) and Graph Neural Networks (GNNs) to learn continuous-time dynamics on graphs. Such a model can tackle continuous-time network dynamics prediction, structured sequence prediction, and node semi-supervised classification on graphs in a unified framework.
    
    \item \mytag{Physical interpretations:} Instead of mapping neural networks through a discrete number of layers, we integrate differential equation systems modeled by GNN over continuous time, which gives physical meaning of continuous-time  dynamics on graphs to GNNs. Our \model can be interpreted as a Continuous-time GNN model or a Graph Neural ODEs.
    \item \mytag{Good performance:} Our experimental results show that our \model  learns continuous-time dynamics on various complex networks accurately, achieves lower error in structured sequence prediction with much fewer parameters than temporal graph neural network models, and outperforms many  GNN models in node semi-supervised classification tasks.
\eit
Our codes and datasets are open-sourced at \url{https://github.com/calvin-zcx/ndcn} .

\section{Related work}
\mytag{Dynamics of complex networks.} Real-world complex systems are usually modeled as complex networks and driven by continuous-time nonlinear dynamics:
the dynamics of brain and human microbial are examined in \cite{gerstner2014neuronal} and \cite{bashan2016universality} respectively; \cite{gao2016universal} 
investigated the resilience dynamics
of complex systems.
\cite{barzel2015constructing} gave a pipeline to construct network dynamics. 
To the best of our knowledge, our \model is the first neural network approach which learns 
continuous-time dynamics on complex networks in a data-driven manner. 

\mytag{Data-driven discovery of differential equations.}
Recently, some data-driven approaches are proposed to learn 
the ordinary/partial differential equations
(ODEs or PDEs),
including sparse regression \cite{mangan2016inferring}, residual network \cite{qin2018data}, feedforward neural network
\cite{raissi2018multistep}, coupled neural networks \cite{raissi2018deep} and so on. \cite{mangan2016inferring} tries to learn continuous-time biological networks dynamics by sparse regression over a large library of interaction terms.
Building the interaction terms are prohibitively for large systems with many interactions due to its combinatorial complexity.
In all, none of them can learn the dynamics on complex systems with more than hundreds of nodes and tens of thousands of interactions.

\mytag{Neural ODEs and Optimal control.}
Inspired by residual network \cite{he2016deep} and ordinary differential equation (ODE) theory \cite{lu2017beyond,ruthotto2018deep}, seminal work neural ODE  model   \cite{chen2018neural} was proposed to re-write residual networks, normalizing flows, and recurrent neural network in a dynamical system way.
See improved Neural ODEs in \cite{dupont2019augmented}.
However, our \model model deals with large differential equations systems. Besides, our model solves different problems, namely learning continuous-time  dynamics on  graphs.
Relationships between back-propagation in deep learning 
and optimal control theory 
are investigated in \cite{han2018mean,benning2019deep}. We formulate our loss function by leveraging the concept of running loss and terminal loss in optimal control. We give novel constraints in optimal control which is modeled by graph neural differential equations systems. Our model solves novel tasks, e.g. learning the  dynamics on complex networks and refer to Sec.\ref{sec:problem}.

\mytag{Graph neural networks and temporal-GNNs.} Graph neural networks (GNNs) \cite{wu2019comprehensive},
e.g., Graph convolution network (GCN) \cite{KipfW17}, attention-based GNN (AGNN) \cite{thekumparampil2018attention}, graph attention networks (GAT) \cite{velivckovic2017graph}, etc.,   
%
achieved very good performance on node semi-supervised learning on graphs.  
However, existing GNNs usually have integer number of 1 or 2 layers \cite{li2018deeper,wu2019comprehensive}. Our \model gives a dynamical system view to GNNs: 
 the continuous depth can be interpreted as continuous physical time, and the outputs of any hidden layer at time $t$ are instantaneous rate of change of network dynamics at that moment.
 By capturing continuous-time network dynamics with real number of depth/time, our model gives very competitive and even better results than above GNNs.
  By combining RNNs or convolution operators with GNNs, temporal-GNNs \cite{yu2017spatio,kazemi2019relational,narayan2018learning,seo2018structured} try to predict next few steps of the  regularly-sampled structured sequences.
 However, these models can not be applied to continuous-time dynamics (observed at arbitrary physical times with different time intervals). Our \model not only predicts the continuous-time network dynamics at an arbitrary time, but also predicts the structured sequences very well with much fewer parameters.


\section{General framework}
\subsection{Problem Definition}
\label{sec:problem}
We can describe the continuous-time dynamics on a graph by a differential equation system:
{
\footnotesize
\begin{equation}
\begin{aligned}
  \frac{d X(t)}{d t}  = f\Big(X, G, W, t\Big),
\end{aligned}
\end{equation}} where $X(t) \in{\mathbb R}^{n \times d} $ represents the state (node feature values) of a dynamic system consisting of $n$ linked nodes at time $t \in [0, \infty) $, and each node is characterized by $d$ dimensional features. $G=(\mathcal{V},\mathcal{E})$ is the network structure capturing how  nodes are interacted with each other. $W(t)$ are parameters which control  how the system evolves over time. $X(0) = X_0$ is the initial states  of this system at time $t=0$. The function $f : {\mathbb R}^{n \times d} \to {\mathbb R}^{n \times d}$ governs the instantaneous rate of change of dynamics on the graph.
In addition, nodes can have various semantic labels encoded by one-hot encoding $Y(X, \Theta) \in \{0, 1\}^{n \times k}$, and $\Theta$ represents the parameters of this classification function. 
The problems we are trying to solve  are:
\bit
    \item \mytag{ (Continuous-time network dynamics prediction) How to predict the continuous-time dynamics $\frac{d X(t)}{d t}$ on a graph  at an arbitrary time?} Mathematically, given a graph $G$ and nodes' states of system $\{\hat{X(t_1)}, \hat{X(t_2)}, ..., \hat{X(t_T)} | 0 \leq t_1 < ... < t_T\}$ where $t_1$ to $t_T$ are arbitrary physical timestamps, can we learn differential equation systems $\frac{d X(t)}{d t} = f(X, G, W, t)$ to  predict continuous-time dynamics $X(t)$ at an arbitrary physical time $t$? The arbitrary physical times mean  that $\{t_1, ..., t_T\}$ are  irregularly sampled with different observational time intervals. When $t > t_T$, we call the prediction task \mytag{extrapolation prediction}, while $t < t_T$ and $t \neq \{t_1, ..., t_T\}$ for \mytag{interpolation prediction}.
    
    \item \mytag{ (Structured sequence prediction)}.    As a special case when $X(t)$ 
    are sampled regularly with the same time intervals $\{\hat{X[1]}, \hat{X[2]}, ..., \hat{X[T]} \}$, the above problem degenerates to a structured sequence learning task with an emphasis on sequential order instead of arbitrary physical times. The goal is to predict next $m$ steps' structured sequence $X[T + 1], ..., X[T + m]$ .
    
    \item \mytag{ (Node semi-supervised classification on graphs) How to predict the unknown nodes'  labels given  features for each node and some labels at one snapshot?} As a special case of above problem with an emphasis on a specific moment, given a graph $G$, one-snapshot features $X$ and some labels $Mask \odot Y$, can we learn a network dynamics  to predict unknown labels $(1-Mask) \odot Y$ by spreading given features and labels on the graph?
\eit
We try to solve above three tasks on learning dynamics on graphs in a unified framework.

\subsection{A Unified Learning Framework}
We formulate our basic framework as follows:
{
\small
\begin{equation}
\begin{aligned}
  & \argmin_{W(t),  \Theta(T)} \mathcal{L} = \int_0^{T} \mathcal{R}\Big (X, G, W, t \Big ) \ d t + \mathcal{S}\Big (Y(X(T), \Theta)\Big )   \\
  & \st \frac{d X(t)}{dt}  = f \Big(X, G, W, t\Big), \ \  X(0)=X_0
\label{equ:optc}
\end{aligned}
\end{equation}}where $\int_0^{T} \mathcal{R}\Big (X, G, W, t \Big ) \ d t$ 
%
 is the { "running" loss} of the continuous-time dynamics on graph from $t=0$ to $t=T$, and $\mathcal{S}(Y(X(T), \Theta)) $
is the { "terminal"  loss} at time $T$.  By integrating $\frac{d X}{d t}= f(X, G, W, t)$ over time $t$ from initial state $X_0$, a.k.a. solving the initial value problem \cite{boyce1992elementary} for this differential equation system, we can get the continuous-time network dynamics $X(t)=X(0) + \int_0^{T} f (X, G, W, \tau ) \ d \tau$ at  arbitrary time moment $t > 0$. 

Such a formulation can be seen as
an optimal control problem so that the goal becomes to learn the best control parameters $W(t)$ for differential equation system $\frac{d X}{d t}= f(X, G, W, t)$ and the best classification parameters $\Theta$ for semantic function $Y(X(t), \Theta)$  by solving above optimization problem. Different from traditional optimal control framework, we model the differential equation systems $\frac{d X}{d t}= f(X, G, W, t)$ by  graph  neural networks. By integrating $\frac{d X}{d t} = f(X, G, W, t)$ over continuous time, namely  $X(t)  = X(0) +  \int_{0}^{t} f \Big(X, G, W, \tau\Big) \ d \tau$, we get our \emph{graph neural ODE} model.
In a dynamical system view, our graph neural ODE  can
be a  time-varying coefficient dynamical system when $W(t)$ changes over time; 
or a constant coefficient dynamical system when $W$ is constant over time for parameter sharing. 
 It's worthwhile to recall that the deep learning methods with $L$ hidden neural layers $f_*$ are $X[L] = f_L \circ ... \circ f_2 \circ f_1 (X[0])$, which are iterated maps \cite{strogatz2018nonlinear} with an integer number of discrete layers and thus can not learn continuous-time dynamics $X(t)$ at arbitrary time. In contrast, our graph neural ODE model $X(t)  = X(0) +  \int_{0}^{t} f \Big(X, G, W, \tau\Big) \ d \tau$ can have continuous layers with a real number $t$ depth corresponding to the continuous-time dynamics on graph $G$. Thus, we can also interpret our graph neural ODE model as a \emph{continuous-time GNN}.

\hide{
An equivalent formulation of Eq.(\ref{equ:optc}) by explicitly solving the initial value problem of differential equation system is:
{\small
\begin{equation}
\begin{aligned}
  & \argmin_{W(t),  \Theta(T)} \mathcal{L} = \int_0^{T} \mathcal{R}\Big (X(t), G, W(t), t \Big ) \ d t + \mathcal{S}\Big (Y(X(T), \Theta)\Big )   \\
  & \st X(t)  = X(0) +  \int_{0}^{t} f \Big(X(\tau), G, W(\tau), \tau\Big) \ d \tau
\label{equ:optc_int}
\end{aligned}
\end{equation}}
}





Moreover, to further increase the express ability 
of our model, 
we  encode the network signal $X(t)$ from the original space to 
$X_h(t)$ in a hidden space, and learn the dynamics in such a hidden space. Then our model becomes:
{
\small
\begin{equation}
\begin{aligned}
& \underset{W(t),  \Theta(T)}{\argmin}
& & \mathcal{L} = \int_0^{T} \mathcal{R}\Big (X, G, W, t \Big ) \ d t + \mathcal{S}\Big (Y(X(T), \Theta)\Big )   \\
& \st
&& X_h(t)  = f_e\Big(X(t), W_e\Big), \ \  X(0)=X_0\\
&&& \frac{d X_h(t)}{dt}  = f \Big(X_h, G, W_h, t\Big), \\
&&& X(t)  = f_d\Big(X_h(t), W_d\Big)
\label{equ:optc_embed}
\end{aligned}
\end{equation}}
where the first constraint transforms $X(t)$  into hidden space $X_h(t)$  through encoding function $f_e$.
The second constraint is the governing dynamics in the hidden space.
The third constraint decodes the hidden signal back to the original space with decoding function $f_d$. The design of $f_e$, $f$, and $f_d$ are flexible to be {\em any deep neural structures}.
We name our graph neural ODE (or continuous-time GNN) model  as {\em Neural Dynamics on Complex Networks} (\mytag{NDCN}).

We solve the initial value problem (i.e., integrating the differential equation systems over time numerically) by  numerical methods (e.g., $1^{st}$-order  Euler method,  high-order method Dormand-Prince DOPRI5 \cite{dormand1996numerical}, etc.). The numerical methods can approximate continuous-time dynamics $X(t)  = X(0) +  \int_{0}^{t} f \Big(X, G, W, \tau\Big) \ d \tau$ at arbitrary time $t$ accurately with guaranteed error. Thus, an equivalent formulation of Eq.(\ref{equ:optc_embed}) by explicitly solving the initial value problem of differential equation system is:
{\small
\begin{equation}
\begin{aligned}
& \underset{W(t),  \Theta(T)}{\argmin}
& & \mathcal{L} = \int_0^{T} \mathcal{R}\Big (X, G, W, t \Big ) \ d t + \mathcal{S}\Big (Y(X(T), \Theta)\Big )   \\
& \st
&& X_h(t)  = f_e\Big(X(t), W_e\Big), \ \ X(0)=X_0\\
&&&  X_h(t) =  X_h(0) + \int_{0}^{t} f \Big(X_h, G, W_h, \tau\Big) d\tau \\
&&& X(t)  = f_d\Big(X_h(t), W_d\Big)
\label{equ:optc_embed_int}
\end{aligned}
\end{equation}}
A large time $t$ corresponds to "deep" neural networks with the physical meaning of a long trajectory of  dynamics on graphs.
In order to learn the learnable parameters $W_*$, we back-propagate the gradients of the loss function w.r.t the control parameters $\frac{\partial \mathcal{L}}{\partial W_*}$ over the numerical integration process backwards in an end-to-end manner, and solve the optimization problem by stochastic gradient descent methods (e.g., Adam \cite{KingmaB14}). 
We will show concrete examples of above general framework in the following three sections.

\section{Learning continuous-time network dynamics}
\label{sec:continuous}


In this section, we investigate if our \model model can learn continuous-time dynamics on graphs for both interpolation prediction and extrapolation prediction. 

\subsection{A Model Instance}
\begin{figure}[ht]
\centering
\includegraphics[width=.5\textwidth]{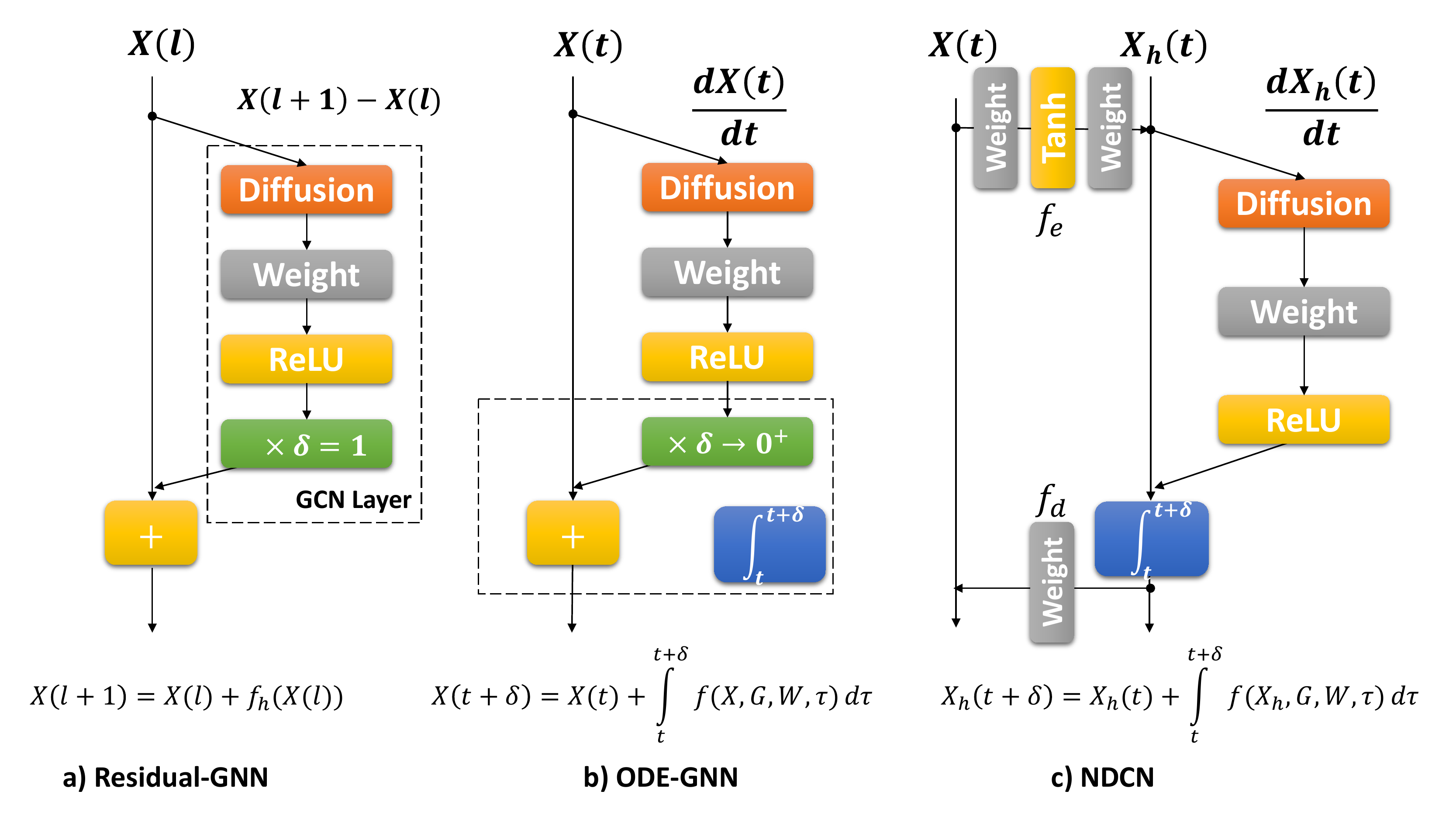}
\vspace{-0.15in}
\caption{ \emph{Illustration of an NDCN instance.} a) a Residual Graph Neural Network (Residual-GNN), b) an Ordinary Differential  Equation GNN model (ODE-GNN), and c) An instance of our  Neural Dynamics on Complex Network (NDCN) model. 
The integer $l$ represents the discrete $l^{th}$ layer and the real number $t$ represents continuous physical time. 
}
\vspace{-1em}
\label{fig:framework}
\end{figure}
We  solve the objective function in (\ref{equ:optc_embed})
with an emphasis on running loss and a fixed graph.
Without the loss of  generality, we use $\ell_1$-norm loss as the running loss $\mathcal{R}$.
More concretely, we adopt two fully connected neural layers with a nonlinear hidden layer as the encoding function $f_e$, a graph convolution neural network (GCN) like structure \cite{KipfW17} but with a simplified  diffusion operator $\Phi$ to model the instantaneous rate of change of  dynamics on graphs in the hidden space, and a linear decoding function $f_d$ for regression tasks in the original signal space. Thus, our model is:

{
\small
\begin{equation}
\begin{aligned}
& \underset{W_*,  b_*}{\argmin}
& & \mathcal{L} = \int_0^{T} \abs { X(t) - \hat{X(t)}} \ dt  \\
& \st
&& X_h(t)  = \tanh \Big(X(t)W_e + b_e\Big)W_0 + b_0, \ \  X_0 \\
&& & \frac{d X_h(t)}{d t}  =  \relu \Big( \Phi X_h(t)W  + b  \Big), \\
&&& X(t)  =  X_h(t)W_d + b_d
\label{equ:optc_run}
\end{aligned}
\end{equation}} where $\hat{X(t)} \in \mathbb{R}^{n \times d}$ is the supervised dynamic information available at time stamp $t$ (in the semi-supervised case the missing information can be padded by 0). The $\abs{\cdot}$ denotes $\ell_1$-norm loss (mean element-wise absolute value difference) between $X(t)$ and $\hat{X(t)}$ at time $t \in [0, T]$.  We illustrate the neural structures of our model in Figure~\ref{fig:framework}c.

We adopt a linear diffusion operator $\Phi = D^{-\frac{1}{2}}(D-A)D^{-\frac{1}{2}} \in \mathbb{R}^{n \times n}$  which is the normalized graph Laplacian where $A\in \mathbb{R}^{n \times n}$ is the adjacency matrix of the network and $D\in \mathbb{R}^{n \times n}$ is the corresponding node degree matrix. The $W \in \mathbb{R}^{d_e \times d_e}$ and $b  \in \mathbb{R}^{n \times d_e}$ are shared parameters 
(namely, the weights and bias of a linear connection layer) over time $t\in [0, T]$. 
The $W_e \in \mathbb{R}^{d \times d_e}$ and $W_0 \in \mathbb{R}^{d_e \times d_e}$ are the matrices in linear layers for encoding, while $W_d \in \mathbb{R}^{d_e \times d}$ are for   decoding.
The $b_e, b_0, b, b_d$ are the biases at the corresponding layer. 
We lean the parameters $W_e, W_0, W, W_d,  b_e, b_0, b, b_d$
from empirical data so that we can learn ${X}$ in a data-driven manner. 
We design the graph neural differential equation system as $\frac{d X(t)}{d t}= \relu(\Phi X(t)W+ b)$ to learn any unknown network dynamics. We can regard $\frac{d X(t)}{d t}$ as a single neural layer at time moment $t$. The $X(t)$ at arbitrary time $t$ is achieved by integrating  $\frac{d X(t)}{d t}$ over time, i.e., $X (t)  = X (0) +  \int_{0}^{t} \relu \Big( \Phi X (\tau)W  + b  \Big) \ d \tau$, leading to a continuous-time graph neural network. 


\subsection{Experiments}

\mytag{Network Dynamics}
\label{sec:dynamics}
We investigate following three continuous-time network dynamics from physics and biology.  
 Let $\overrightarrow{x_i(t)} \in \mathbb{R}^{d\times 1}$ be   $d$ dimensional features of node $i$ at time $t$ and thus $ X(t) = [\dots, \overrightarrow{x_i(t)}, \dots]^T \in \mathbb{R}^{n \times d}$. We show their differential equation systems in vector form  for clarity and implement them in matrix form:
\bit
    \item The heat diffusion dynamics $\frac{d\overrightarrow{x_i(t)}}{dt} =  -k_{i,j} \sum\nolimits_{j=1}^{n} A_{i,j} (\overrightarrow{x_i} - \overrightarrow{x_j})$  
     governed by Newton's law of cooling \cite{luikov2012analytical}, 
    which states that the rate of heat change of node $i$ is proportional to the difference of the temperature between node $i$ and its neighbors with heat capacity matrix $A$. 


    \item The mutualistic interaction dynamics among species in ecology, governed by equation $\frac{d \overrightarrow{x_i(t)}}{d t} = b_i + \overrightarrow{x_i} (1 - \frac{\overrightarrow{x_i}}{k_i})(\frac{\overrightarrow{x_i}}{c_i} -1) + \sum_{j=1}^{n} A_{i,j} \frac{\overrightarrow{x_i} \overrightarrow{x_j}}{d_i + e_i \overrightarrow{x_i} + h_j \overrightarrow{x_j}}$ 
    ({For brevity, the operations between vectors are element-wise}).
    \hide{
    \scriptsize
    \begin{equation} \label{equ:mut}
        \begin{aligned} 
              \frac{d \overrightarrow{x_i(t)}}{d t} = b_i + \overrightarrow{x_i} (1 - \frac{\overrightarrow{x_i}}{k_i})(\frac{\overrightarrow{x_i}}{c_i} -1) + \sum_{j=1}^{n} A_{i,j} \frac{\overrightarrow{x_i} \overrightarrow{x_j}}{d_i + e_i \overrightarrow{x_i} + h_j \overrightarrow{x_j}}.
        \end{aligned}
    \end{equation}}
    The mutualistic differential equation systems \cite{gao2016universal} capture the abundance $\vec{x_i(t)}$ of species $i$, consisting of incoming migration term $b_i$, logistic growth with population capacity $k_i$ \cite{zang2018power} and Allee effect \cite{allee1949principles} with cold-start threshold $c_i$, and mutualistic interaction term with interaction network $A$. 
    
    \item The gene regulatory dynamics governed by Michaelis-Menten equation $\frac{d \overrightarrow{x_i(t)}}{d t} =  - b_i \overrightarrow{x_i}^{f} + \sum\nolimits_{j=1}^{n} A_{i,j}\frac{\overrightarrow{x_j}^{h}}{\overrightarrow{x_j}^{h} + 1}$
    \hide{\small
    \begin{equation} \label{equ:gene}
        \begin{aligned} 
              \frac{d \overrightarrow{x_i(t)}}{d t} =  - b_i \vec{x_i}^{f} + \sum\nolimits_{j=1}^{n} A_{i,j}\frac{\overrightarrow{x_j}^{h}}{\overrightarrow{x_j}^{h} + 1},
        \end{aligned}
    \end{equation}}
    where the first term models degradation when $f=1$ or dimerization when $f=2$, and the second term captures genetic activation tuned by the  Hill coefficient $h$  \cite{alon2006introduction,gao2016universal}.
\eit

\begin{figure*}[t]
\centering
\includegraphics[width=.65\textwidth]{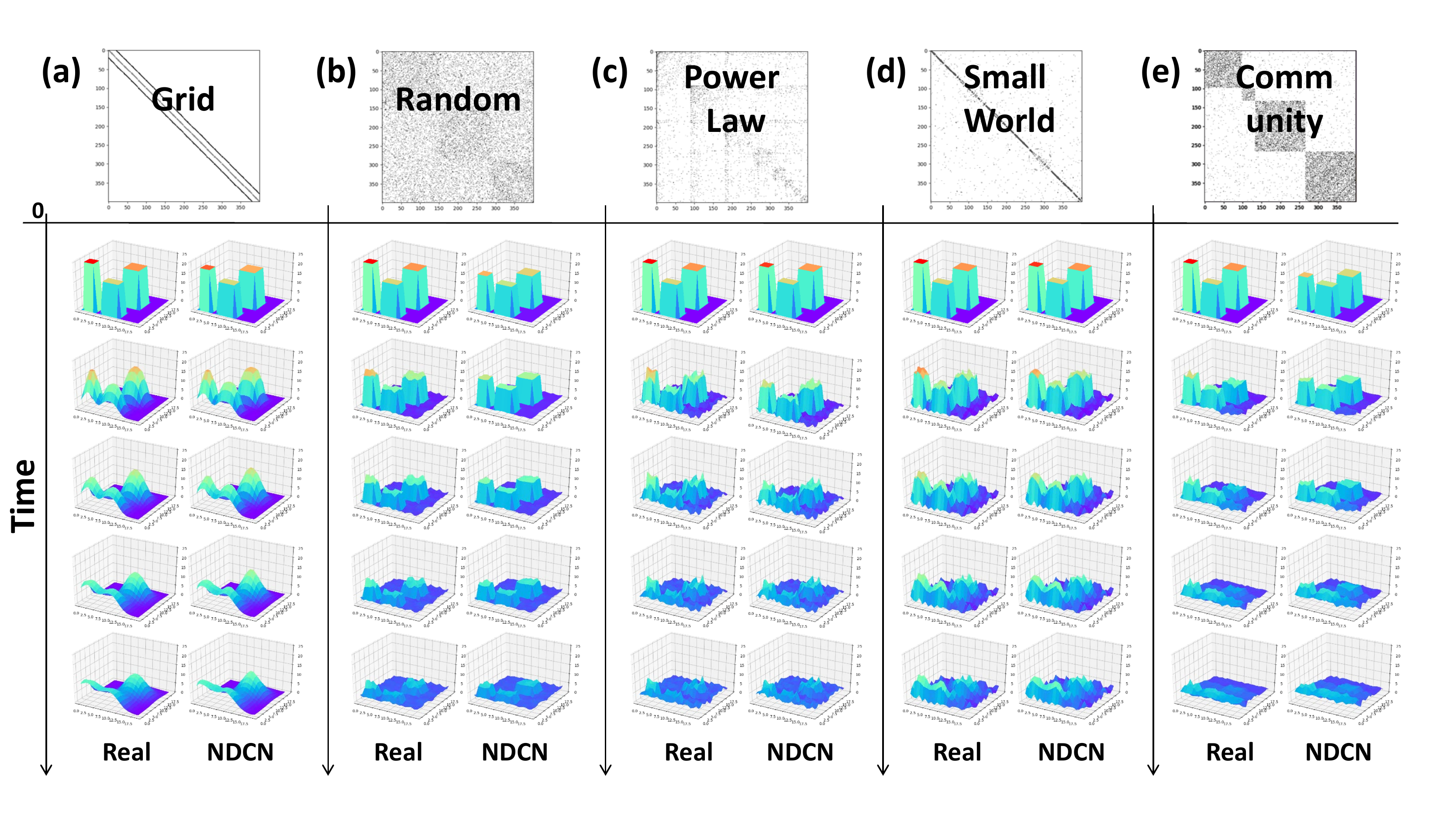}
\caption{ \emph{Heat diffusion on different networks. Our \model fits the dynamics on different networks accurately. } 
Each of the five vertical panels (a)-(e) represents the dynamics on one network over physical time. For each network dynamics, we illustrate the sampled ground truth dynamics (left) and the dynamics generated by our \model (right) from top to bottom following the direction of time. Refer to Appendix~\ref{appendix:animation} for the animations.
}
\vspace{-1em}
\label{fig:heat}
\end{figure*}

\mytag{Complex Networks.} We consider following networks:
(a) Grid network, where each node is connected with 8 neighbors (as shown in Fig.~\ref{fig:heat}(a)) 
;
(b) Random network, generated by Erd{\'o}s and R{\'e}nyi model \cite{erdds1959random} (as shown in Fig.~\ref{fig:heat}(b));
(c) Power-law network, generated by Albert-Barab{\'a}si model \cite{barabasi1999emergence} (as shown in Fig.~\ref{fig:heat}(c));
(d) Small-world network, generated by Watts-Strogatz model \cite{watts1998collective} (as shown in Fig.~\ref{fig:heat}(d));
and (e) Community network, generated by random partition model \cite{fortunato2010community} (as shown in Fig.~\ref{fig:heat}(e)).

\mytag{Visualization.}
\label{sec:viz}
To visualize dynamics on complex networks over time is not trivial. 
We first generate a network with $n$ nodes by aforementioned network models. The nodes are re-ordered according to the community detection method by Newman \cite{newman2010networks} and each node has a unique label from $1$ to $n$. We layout these nodes on a 2-dimensional $\sqrt{n} \times \sqrt{n}$ grid and each grid point $(r,c) \in \mathbb{N}^2$ represents the $i^{th}$ node where $i=r\sqrt{n}+c+1$. Thus, nodes' states $X(t) \in \mathbb{R}^{n\times d}$ at time $t$ when $d=1$ can be visualized as a scalar field function $X: \mathbb{N}^2 \to \mathbb{R}$ over the grid. 
Please refer to Appendix~\ref{appendix:animation} for the animations of these  dynamics on different complex networks over time.

\mytag{Baselines.} To the best of  our knowledge,  there are no baselines for learning continuous-time  dynamics on complex networks, and thus we compare the ablation models of \model for this task.  By investigating ablation models
we show that our \model is a minimum model for this task.
We keep the loss function  same and construct  following baselines:
\bit
    \item The model without encoding  $f_e$ and $f_d$ and thus no hidden space: 
    $ \frac{d X(t)}{d t}  =  \relu ( \Phi X(t)W + b) $
    , namely ordinary differential equation GNN model (ODE-GNN),  which learns the dynamics in the original signal space $X(t)$ as shown in Fig.~\ref{fig:framework}b;
    
    \item The model without  graph diffusion operator $\Phi$: 
    $ \frac{d X_h(t)}{d t}  =  \relu (X_h(t)W + b)$, i.e., an neural ODE model \cite{chen2018neural}, which can be thought as a continuous-time version of forward residual  neural network (See Fig.~\ref{fig:framework}a and Fig.~\ref{fig:framework}b for the difference between residual network and ODE network). 
    
    \item The model without control parameters, namely weight layer $W$: 
    $ \frac{d X_h(t)}{d t}  =  \relu (\Phi X_h(t)) $
    which has no linear connection layer  between $t$ and $t +  dt$ (where $dt \to 0$) and thus indicating a determined dynamics to spread signals (See Fig.~\ref{fig:framework}c without a weight layer).
\eit

\begin{figure*}[!ht]
\centering
\includegraphics[width=.7\textwidth]{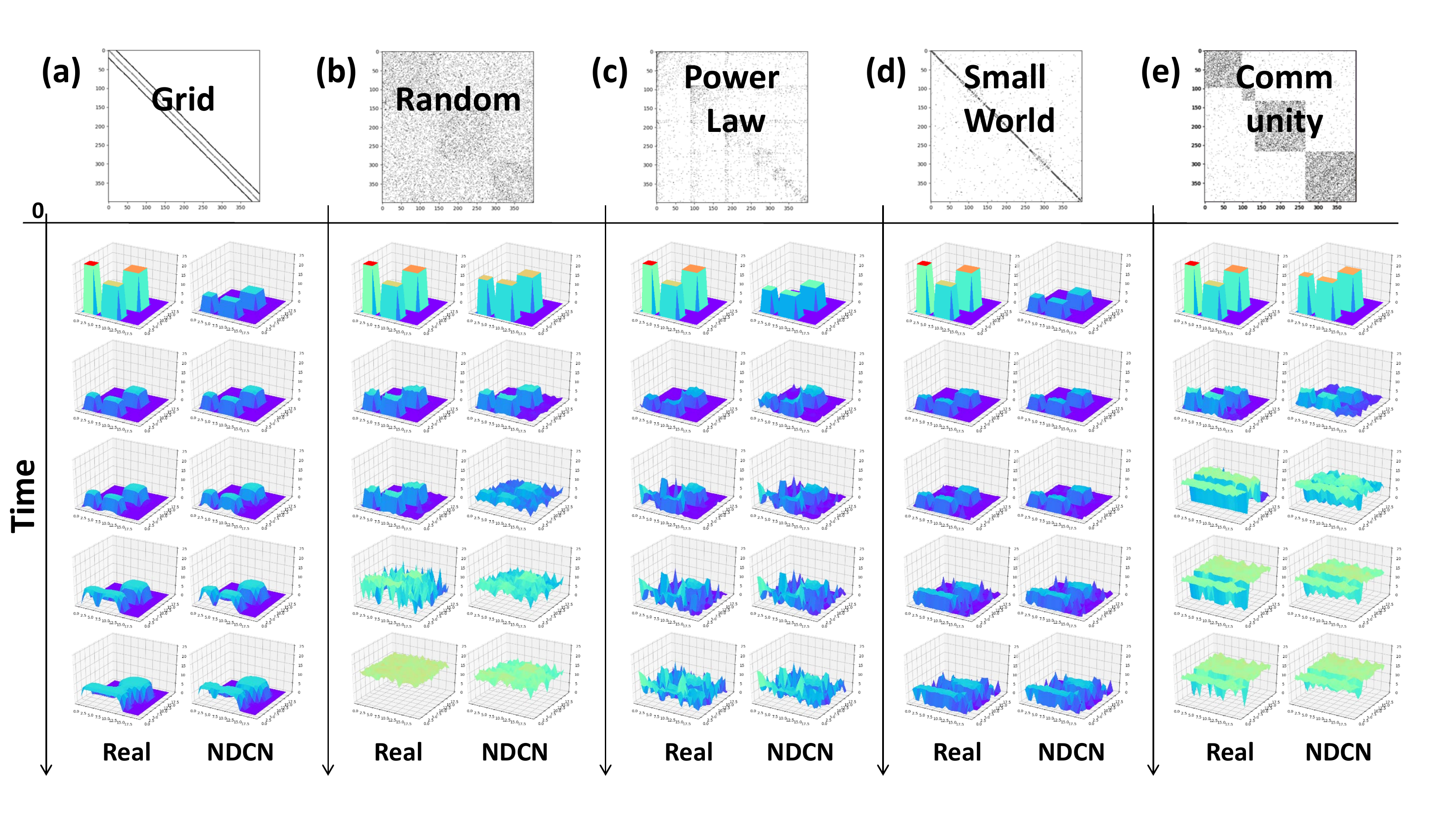}
\vspace{-0.15in}
\caption{ \emph{Biological mutualistic interaction on different networks. Our \model fits the dynamics on different networks accurately. Refer to Appendix~\ref{appendix:animation} for the animations.} 
}\label{fig:mutualistic}
\vspace{-0.5em}
\end{figure*}

\begin{figure*}[!ht]
\centering
\includegraphics[width=.7\textwidth]{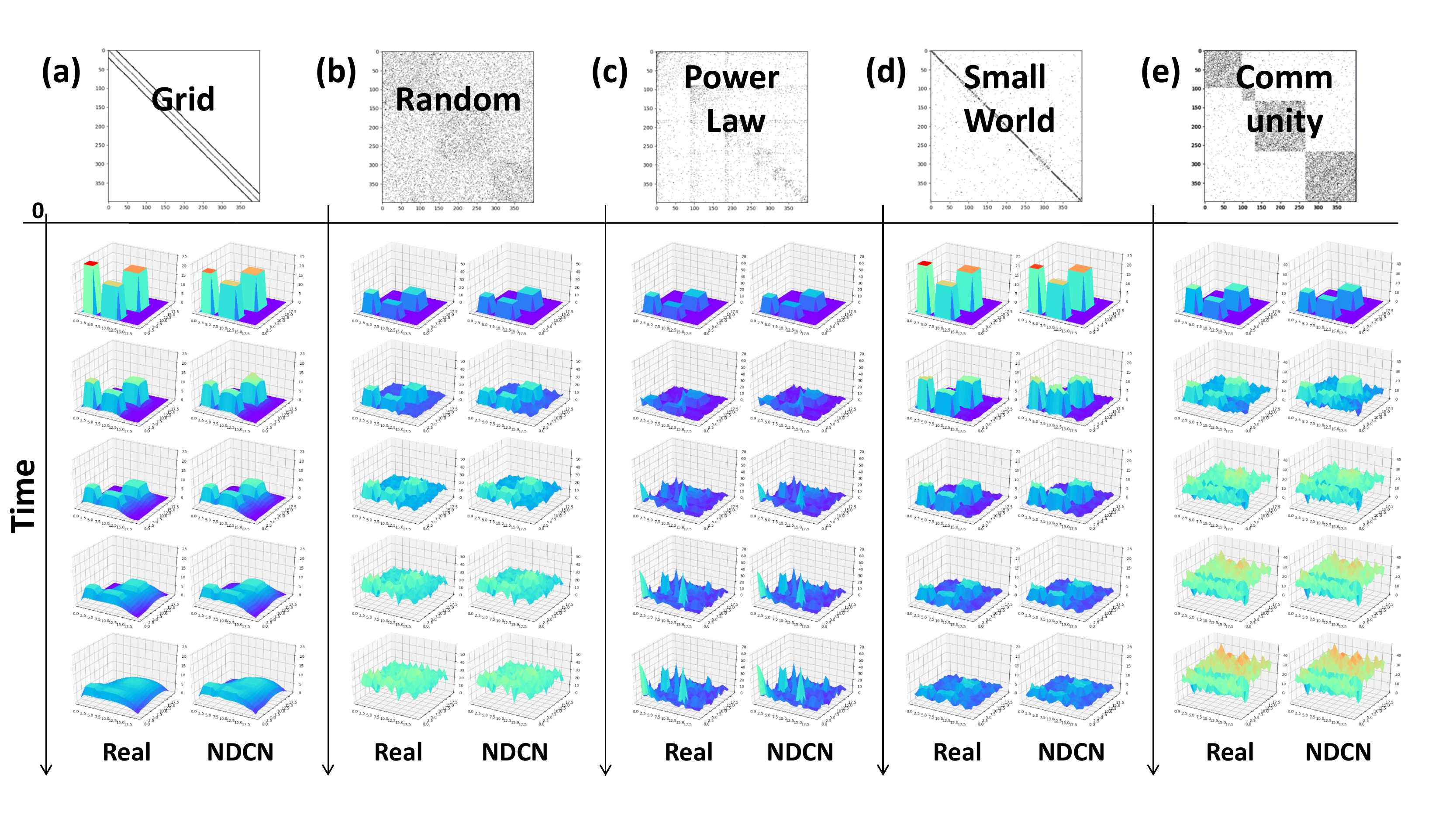}
\vspace{-0.15in}
\caption{ \emph{Gene regulation dynamics on different networks. Our \model fits the dynamics on different networks accurately. Refer to Appendix~\ref{appendix:animation} for the animations.} 
}
\vspace{-0.5em}
\label{fig:gene}
\end{figure*}

\mytag{Experimental setup.}  
We  generate underlying networks with $400$ nodes by network models in Sec.\ref{sec:viz}  and the illustrations are shown
in Fig.~\ref{fig:heat},\ref{fig:mutualistic} and \ref{fig:gene}.
We set the initial value $X(0)$ the same
for all the experiments
and thus different dynamics are only due to their different dynamic rules   and underlying networks (See Appendix~\ref{appendix:animation}).

We irregularly sample 120 snapshots of the continuous-time dynamics $\{\hat{X(t_1)}, ..., \hat{X(t_{120})}  | 0 \leq  t_1 < ... < t_{120} \leq T \}$ where the time intervals between $t_1$, ..., $t_{120}$ are different.  We randomly choose 80 snapshots from $\hat{X(t_1)}$ to $\hat{X(t_{100})}$ for training, the left 20 snapshots from $\hat{X(t_1)}$ to $\hat{X(t_{100})}$ for testing the interpolation prediction task. We use the 20 snapshots from $\hat{X(t_{101})}$ to $\hat{X(t_{120})}$ for testing the extrapolation prediction task.



We use Dormand-Prince method \cite{dormand1996numerical} to get the ground truth dynamics, and use Euler method in the forward process of our \model 
(More configurations in Appendix~\ref{appendix:conf}). We evaluate the results by $\ell_1$ loss
and normalized $\ell_1$ loss (normalized by the mean element-wise value of $\hat{X(t)}$), and they lead to the same conclusion (We report normalized $\ell_1$ loss here and see Appendix~\ref{appendix:abs} for $\ell_1$ loss). Results are the mean and standard deviation of the loss over $20$ independent runs for $3$ dynamic laws on $5$ different networks by each method.


\begin{table}[th]
\vspace{-.5em}
\centering
\caption{ \mytag{Extrapolation of continuous-time network dynamics.} Our \model predicts different continuous-time network dynamics accurately. Each result is the normalized $\ell_1$ error with standard deviation (in percentage $\%$) from $20$ runs for 3 dynamics on 5 networks by each method.
 }
\label{table:fit_dyn}
\tiny
\begin{tabular}{l l l l l l l}
\hline
  & & Grid & Random & Power Law & Small World & Community \\ \hline
\multirow{4}{*}{\makecell{Heat \\ Diffusion}} 
 & No-Encode  &  $29.9 \pm 7.3$ & $27.8 \pm 5.1$ & $24.9 \pm 5.2$& $24.8 \pm 3.2$& $30.2 \pm 4.4$\\ 
 & No-Graph  & $30.5 \pm 1.7$ &$5.8\pm 1.3$ &$6.8 \pm 0.5$ & $ 10.7\pm 0.6$& $ 24.3\pm 3.0 $\\
 & No-Control  &  $ 73.4\pm 14.4 $& $28.2\pm 4.0$& $25.2\pm 4.3$& $ 30.8\pm 4.7$&$37.1 \pm 3.7$\\
 & \textbf{\model}  &  $\mathbf{ 4.1\pm 1.2}$ & $\mathbf{ 4.3 \pm 1.6}$ & $\mathbf{ 4.9 \pm 0.5}$ & $\mathbf{ 2.5 \pm 0.4}$ & $\mathbf{4.8 \pm 1.0}$\\ \hline 
\multirow{4}{*}{\makecell{Mutualistic \\ Interaction}} 
 & No-Encode  &  $ 45.3\pm 3.7$ & $ 9.1 \pm 2.9$ & $ 29.9 \pm 8.8$ & $54.5 \pm 3.6$ & $14.5 \pm 5.0$\\
 & No-Graph  &  $ 56.4 \pm 1.1$ & $ 6.7 \pm 2.8$ & $14.8 \pm 6.3$ & $54.5 \pm 1.0$ & $ 9.5 \pm 1.5$\\
 & No-Control  &  $140.7 \pm 13.0$ & $10.8 \pm 4.3$ & $106.2 \pm 42.6$ & $115.8 \pm 12.9$ & $16.9 \pm 3.1$\\
 & \textbf{\model}  &  $\mathbf { 26.7 \pm 4.7 }$ & $\mathbf{ 3.8 \pm 1.8}$ & $\mathbf{ 7.4 \pm 2.6}$ & $\mathbf{ 14.4 \pm 3.3}$ & $\mathbf{ 3.6 \pm 1.5}$\\ \hline 
\multirow{4}{*}{\makecell{Gene \\ Regulation}} 
 & No-Encode  &  $ 31.7 \pm 14.1$ & $ 17.5 \pm 13.0$ & $33.7 \pm 9.9$ & $25.5 \pm 7.0$ & $26.3 \pm 10.4$\\
 & No-Graph  &  $ 13.3 \pm 0.9$ & $12.2 \pm 0.2$ & $43.7 \pm 0.3$ & $15.4 \pm 0.3$ & $19.6 \pm 0.5$\\
 & No-Control  &  $65.2 \pm 14.2$ & $68.2 \pm 6.6$ & $70.3 \pm 7.7$ & $58.6 \pm 17.4$ & $64.2 \pm 7.0$\\
 & \textbf{\model}  & $\mathbf{  16.0 \pm 7.2}$ & $\mathbf{ 1.8 \pm 0.5}$ & $\mathbf{ 3.6 \pm 0.9}$ & $\mathbf{ 4.3 \pm 0.9}$ & $\mathbf{ 2.5 \pm 0.6}$\\ \hline
\end{tabular}
\end{table}

\begin{table}[th]
\vspace{-0.5em}
\centering
\caption{ \mytag{Interpolation of continuous-time network dynamics.} Our \model predicts different continuous-time network dynamics accurately. Each result is the normalized $\ell_1$ error with standard deviation (in percentage $\%$) from $20$ runs for 3 dynamics on 5 networks by each method.
 }
\label{table:irr_inter}
\tiny
\begin{tabular}{l l l l l l l}
\hline
  & & Grid & Random & Power Law & Small World & Community \\ \hline
\multirow{4}{*}{\makecell{Heat \\ Diffusion}} 
 & No-Encode  &  $32.0 \pm 12.7$ & $26.7 \pm 4.4$ & $25.7 \pm 3.8$ & $27.9 \pm 7.3$ & $35.0 \pm 6.3$\\ 
 & No-Graph  & $41.9 \pm 1.8$ & $9.4\pm 0.6$ & $18.2 \pm 1.5$ & $ 25.0 \pm 2.1$ & $ 25.0\pm 1.4 $\\
 & No-Control  &  $ 56.8\pm 2.8 $& $32.2\pm 7.0$& $33.5\pm 5.7$& $ 40.4\pm 3.4$ & $39.1 \pm 4.5$\\
 & \textbf{\model}  &  $\mathbf{ 3.2\pm 0.6}$ & $\mathbf{ 3.2 \pm 0.4}$ & $\mathbf{ 5.6 \pm 0.6}$ & $\mathbf{ 3.4 \pm 0.4}$ & $\mathbf{4.3 \pm 0.5}$\\ \hline 
\multirow{4}{*}{\makecell{Mutualistic \\ Interaction}} 
 & No-Encode  &  $ 28.9\pm 2.0$ & $ 19.9 \pm 6.5$ & $ 34.5 \pm 13.4$ & $27.6 \pm 2.6$ & $25.5 \pm 8.7$\\
 & No-Graph  &  $ 28.7 \pm 4.5$ & $ 7.8 \pm 2.4$ & $23.2 \pm 4.2$ & $26.9 \pm 3.8$ & $ 14.1 \pm 2.4$\\
 & No-Control  &  $72.2 \pm 4.1$ & $22.5 \pm 10.2$ & $63.8 \pm 3.9$ & $67.9 \pm 2.9$ & $33.9 \pm 12.3$\\
 & \textbf{\model}  &  $\mathbf { 7.6 \pm 1.1 }$ & $\mathbf{ 6.6 \pm 2.4}$ & $\mathbf{ 6.5 \pm 1.3}$ & $\mathbf{ 4.7 \pm 0.7}$ & $\mathbf{ 7.9 \pm 2.9}$\\ \hline 
\multirow{4}{*}{\makecell{Gene \\ Regulation}} 
 & No-Encode  &  $ 39.2 \pm 13.0$ & $ 14.5 \pm 12.4$ & $33.6 \pm 10.1$ & $27.7 \pm 9.4$ & $21.2 \pm 10.4$\\
 & No-Graph  &  $ 25.2 \pm 2.3$ & $11.9 \pm 0.2$ & $39.4 \pm 1.3$ & $15.7 \pm 0.7$ & $18.9 \pm 0.3$\\
 & No-Control  &  $66.9 \pm 8.8$ & $31.7 \pm 5.2$ & $40.3 \pm 6.6$ & $49.0 \pm 8.0$ & $35.5 \pm 5.3$\\
 & \textbf{\model}  & $\mathbf{  5.8 \pm 1.0}$ & $\mathbf{ 1.5 \pm 0.6}$ & $\mathbf{ 2.9 \pm 0.5}$ & $\mathbf{ 4.2 \pm 0.9}$ & $\mathbf{ 2.3 \pm 0.6}$\\ \hline
\end{tabular}
\vspace{-0.5em}
\end{table}

 \mytag{Results.} We visualize the ground-truth and learned dynamics in Fig.~\ref{fig:heat},\ref{fig:mutualistic} and \ref{fig:gene}, and please see the animations of these network dynamics in  Appendix~\ref{appendix:animation}. 
We find  that one dynamic law may behave quite different on different networks: heat dynamics may gradually die out to be stable but follow different dynamic patterns in Fig.~\ref{fig:heat}. Gene dynamics are asymptotically stable  on grid in Fig.~\ref{fig:gene}a but unstable on random networks in Fig.~\ref{fig:gene}b or community  networks in Fig.~\ref{fig:gene}e. Both gene regulation dynamics in Fig.~\ref{fig:gene}c and biological mutualistic dynamics in Fig.~\ref{fig:mutualistic}c  show very bursty patterns on power-law networks. However, visually speaking, our \model learns all these different network dynamics very well.

The quantitative results of extrapolation and interpolation prediction are summarized in Table~\ref{table:fit_dyn} and Table~\ref{table:irr_inter} respectively.
We observe that our \model captures different dynamics on various complex networks accurately and outperforms all the continuous-time baselines
by a large margin, indicating that our \model potentially serves as a  minimum model in learning continuous-time dynamics on complex networks.

\hide{
We also examine other experimental settings in $\Dot{X}= f(X, G, W, t)$: (a) other activation functions like hyperbolic tangent function $\tanh$; (b) graph operator including Laplacian without being normalized, convolution operator adopted in graph convolution neural network \cite{KipfW17}, etc.; (c) different numerical integration method for initial value problem like Euler's method \cite{boyce1992elementary}, etc.;  they all get worse results than our model and we do not report them here for brevity (Appendix~\ref{appendix:results}).
}

\section{Learning Structured Sequences}
\label{sec:regular}
Besides, we can easily use our \model model for learning regularly-sampled structured sequence of network dynamics  and predict future steps by using $1^{st}$-order  Euler method with time step $1$ in the forward integration process. 

\mytag{Experimental setup.}
We regularly sample 100 snapshots of the continuous-time network dynamics discussed in the last section with same time intervals from $0$ to $T$, and denote these structured sequence as $\{\hat{X[1]}, ..., \hat{X[{100}]} \}$.
We use first   80 snapshots $\hat{X[1]}, ..., \hat{X[{80}]}$ for training and the left 20 snapshots $\hat{X[{81}]}, ..., \hat{X[{100}]}$ for testing extrapolation prediction task. The temporal-GNN models are usually used for next few step prediction and can not be used  for the interpolation task (say, to predict $X[{1.23}]$) directly. We use 5 and 10 for hidden dimension of GCN and RNN models respectively. We use $1^{st}$-order  Euler method with time step $1$ in the forward integration process.  Other settings are the same as previous continuous-time dynamics experiment.

\mytag{Baselines.}
We compare our model with the temporal-GNN models  which are usually combinations of RNN models and GNN models \cite{kazemi2019relational,narayan2018learning,seo2018structured}. 
We use GCN \cite{KipfW17}  as a graph structure extractor and use LSTM/GRU/RNN \cite{lipton2015critical} to  learn the temporal relationships between ordered structured sequences. 
We keep the loss function same and construct following baselines:
    
    
\bit
    \item LSTM-GNN: the temporal-GNN with LSTM cell $X[t+1] = LSTM(GCN(X[t], G))$: 
\begin{equation}
\footnotesize
\begin{aligned} 
x_t &= ReLU(\Phi X[t] W_e+b_e) \\
i_t &= \sigma(W_{ii} x_t + b_{ii} + W_{hi} h_{t-1} + b_{hi}) \\
f_t &= \sigma(W_{if} x_t + b_{if} + W_{hf} h_{t-1} + b_{hf}) \\
g_t &= \tanh(W_{ig} x_t + b_{ig} + W_{hg} h_{t-1} + b_{hg}) \\
o_t &= \sigma(W_{io} x_t + b_{io} + W_{ho} h_{t-1} + b_{ho}) \\
c_t &= f_t * c_{t-1} + i_t * g_t \\
h_t &= o_t * \tanh(c_t) \\
\hat{X[t+1]} &= W_d * h_t + b_d\\
\end{aligned}
\end{equation}

    \item GRU-GNN: the temporal-GNN with GRU cell $X[t+1]  = GRU(GCN(X[t], G))$.:  
\begin{equation}
\footnotesize
\begin{aligned} 
x_t &= ReLU(\Phi X[t] W_e+b_e) \\
r_t &= \sigma(W_{ir} x_t + b_{ir} + W_{hr} h_{t-1} + b_{hr}) \\
z_t &= \sigma(W_{iz} x_t + b_{iz} + W_{hz} h_{t-1} + b_{hz}) \\
n_t &= \tanh(W_{in} x_t + b_{in} + r * (W_{hn} h_{t-1} + b_{hn})) \\
h_t &= (1 - z_t) * n_t + z_t * h_{t-1}\\
\hat{X[t+1]} &= W_d * h_t + b_d\\
\end{aligned}
\end{equation}
    
    \item RNN-GNN: the temporal-GNN with RNN cell $X[t+1]  = RNN(GCN(X[t], G))$:
\begin{equation}
\footnotesize
\begin{aligned} 
x_t &= ReLU(\Phi X[t] W_e+b_e) \\   
h_t &= \tanh(w_{ih} x_t + b_{ih}  +  w_{hh} h_{t-1} + b_{hh})\\
\hat{X[t+1]} &= W_d * h_t + b_d\\
\end{aligned}
\end{equation}
\eit

\mytag{Results.}
We summarize the results of the extrapolation prediction of regularly-sampled structured sequence in Table~\ref{table:equal_ex}.  The GRU-GNN model works well in mutualistic dynamics on random network and community network. Our \model predicts different dynamics on these complex networks accurately and outperforms the baselines in almost all the settings. What's more, our model predicts the structured sequences in a much more succinct way  with much fewer parameters.
The learnable parameters of RNN-GNN, GRU-GNN, LSTM-GNN are $24530$, $64770$, and $84890$ respectively. In contrast, our \model  has only \mytag{901} parameters, accounting for  $3.7\%$, $1.4\%$ , $1.1\%$ of above three temporal-GNN models respectively. 

\begin{table}[th]
\vspace{-.5em}
\centering
\caption{ \mytag{Extrapolation prediction for the regularly-sampled structured sequence.} Our \model predicts different structured sequences accurately. Each result is the normalized $\ell_1$ error with standard deviation (in percentage $\%$) from $20$ runs for 3 dynamics on 5 networks by each method.
 }
\label{table:equal_ex}
\tiny
\begin{tabular}{l l l l l l l}
\hline
  & & Grid & Random & Power Law & Small World & Community \\ \hline
\multirow{4}{*}{\makecell{Heat \\ Diffusion}} 
 & LSTM-GNN &  $12.8 \pm 2.1$ & $21.6 \pm 7.7$ & $12.4 \pm 5.1$ & $11.6 \pm 2.2$ & $13.5 \pm 4.2$\\ 
 & GRU-GNN  & $11.2 \pm 2.2$ & $9.1\pm 2.3$ & $ 8.8 \pm 1.3$ & $ 9.3 \pm 1.7$ & $ 7.9\pm 0.8 $\\
 & RNN-GNN  &  $ 18.8\pm 5.9 $& $25.0\pm 5.6$ & $18.9\pm 6.5$& $ 21.8\pm 3.8$ & $16.1 \pm 0.0$\\
 & \textbf{\model}  &  $\mathbf{ 4.3\pm 0.7}$ & $\mathbf{ 4.7 \pm 1.7}$ & $\mathbf{ 5.4 \pm 0.4}$ & $\mathbf{ 2.7 \pm 0.4}$ & $\mathbf{5.3 \pm 0.7}$\\ \hline 
\multirow{4}{*}{\makecell{Mutualistic \\ Interaction}} 
 & LSTM-GNN  &  $ 51.4\pm 3.3$ & $ 24.2 \pm 24.2$ & $ 27.0 \pm 7.1$ & $58.2 \pm 2.4$ & $25.0 \pm 22.3$\\
 & GRU-GNN  &  $ 49.8 \pm 4.1$ & $ \mathbf{1.0 \pm 3.6}$ & $12.2 \pm 0.8$ & $51.1 \pm 4.7$ & $ \mathbf{3.7 \pm 4.0}$\\
 & RNN-GNN  &  $56.6 \pm 0.1$ & $8.4 \pm 11.3$ & $12.0 \pm 0.4$ & $57.4 \pm 1.9$ & $8.2 \pm 6.4$\\
 & \textbf{\model}  &  $\mathbf { 29.8 \pm 1.6 }$ & $\mathbf{ 4.7 \pm 1.1}$ & $\mathbf{ 11.2 \pm 5.0}$ & $\mathbf{ 15.9 \pm 2.2}$ & $\mathbf{ 3.8 \pm 0.9}$\\ \hline 
\multirow{4}{*}{\makecell{Gene \\ Regulation}} 
 & LSTM-GNN  &  $ 27.7 \pm 3.2$ & $ 67.3 \pm 14.2$ & $38.8 \pm 12.7$ & $13.1 \pm 2.0$ & $53.1 \pm 16.4$\\
 & GRU-GNN  &  $ 24.2 \pm 2.8$ & $50.9 \pm 6.4$ & $35.1 \pm 15.1$ & $11.1 \pm 1.8$ & $46.2 \pm 7.6$\\
 & RNN-GNN  &  $28.0 \pm 6.8$ & $56.5 \pm 5.7$ & $42.0 \pm 12.8$ & $14.0 \pm 5.3$ & $46.5 \pm 3.5$\\
 & \textbf{\model}  & $\mathbf{  18.6 \pm 9.9}$ & $\mathbf{ 2.4 \pm 0.9}$ & $\mathbf{ 4.1 \pm 1.4}$ & $\mathbf{ 5.5 \pm 0.8}$ & $\mathbf{ 2.9 \pm 0.5}$\\ \hline
\end{tabular}
\vspace{-1.5em}
\end{table}

\vspace{-.5em}
\section{Node semi-supervised classification}
\label{sec:semantic}
\begin{figure*}[!ht]
\centering
\includegraphics[width=.65\textwidth]{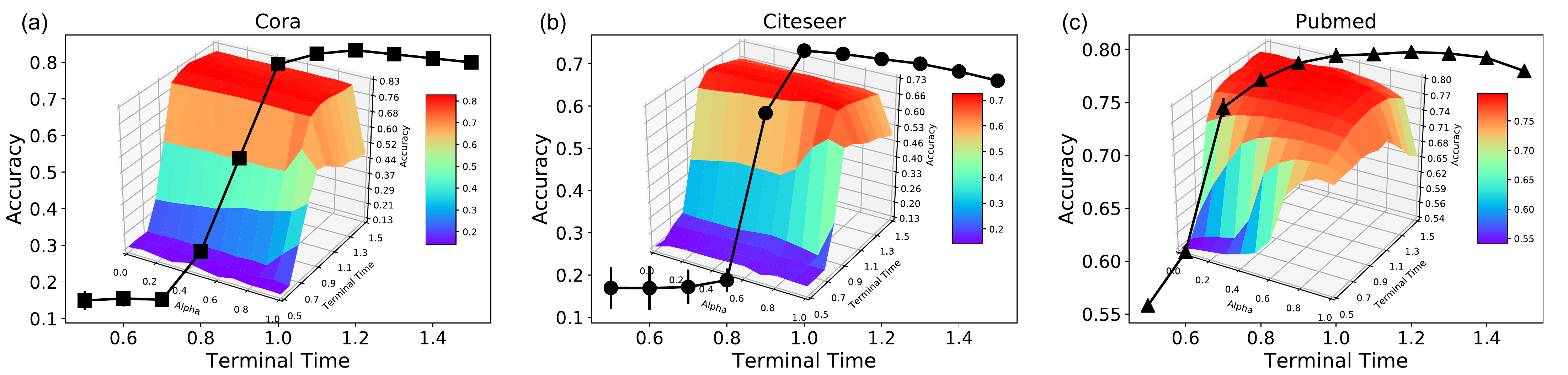}
\vspace{-0.15in}
\caption{ \emph{Our  \model  model captures continuous-time dynamics.}  Mean classification accuracy of $100$ runs over terminal time when given a specific $\alpha$. Insets are the accuracy over the two-dimensional space of terminal time and $\alpha$. The best accuracy is achieved at a real-number time/depth. }
\label{fig:gnn_time}
\vspace{-0.05in}
\end{figure*}
We investigate the third question, i.e., { how to predict the semantic labels of each node given semi-supervised information?}  
Various graph neural networks (GNN) \cite{wu2019comprehensive} achieve very good performance in 
graph semi-supervised classification task \cite{YangCS16,KipfW17}. 
Existing GNNs usually adopt an integer number of 1 or 2 hidden layers \cite{KipfW17,velivckovic2017graph}. Our framework follows the perspective of a dynamical system: by modeling the continuous-time  dynamics to spread nodes' features and given labels on graphs, we predict the unknown labels in analogy to predicting the label dynamics at some  time $T$.


\subsection{A Model Instance}
Following the same framework as in Section 3, we propose a simple model with the terminal semantic loss $\mathcal{S}(Y(T)) $ modeled by the cross-entropy loss for classification task:
%
{
\footnotesize
\begin{equation}
\begin{aligned}
& \underset{W_e, b_e,  W_d, b_d}{\argmin}
& & \mathcal{L} = \int_0^{T} \mathcal{R} (t) \ d t - \sum_{i=1}^{n}\sum_{k=1}^{c} \hat{Y}_{i,k} \log Y_{i, k}(T) \\
& \st
&& X_h(0)  = \tanh \Big(X(0)W_e + b_e\Big)\\ 
&& & \frac{d X_h(t) }{d t} = \relu \Big( \Phi X_h(t) \Big) \\
&& & Y(T)  =  \softmax ( X_h(T)W_d + b_d)
\label{equ:optc_terminal}
\end{aligned}
\end{equation}}
where the $\hat{Y} \in \mathbb{R}^{n \times c}$ is the supervised information of nodes' labels and $Y(T) \in \mathbb{R}^{n \times c}$  is the label dynamics of nodes at time $T \in \mathbb{R}$ whose element $Y_{i, k}(T)$ denotes the probability of the node $i = {1, \dots, n}$ with label $k = {1, \dots, c}$. We use differential equation system  $\frac {d X(t)}{d t}= \relu(\Phi X(t))$ together with encoding and decoding layers to spread  the features and given labels of nodes on graph over continuous time $[0, T]$, i.e., $X_h(T)  = X_h(0) +  \int_{0}^{T} \relu \Big( \Phi X_h(t)\Big) \ d t$.


Compared with the model in Eq.\ref{equ:optc_run}, we only have a one-shot supervised information $\hat{Y}$ given nodes' features $X$. 
Thus, we model the running loss $\int_0^{T} \mathcal{R} (t) \ d t$ as the $\ell_2$-norm regularizer of the learnable parameters 
%
$\int_0^{T} \mathcal{R} (t) \ d t =  \lambda(\abs{W_e}^2_2 + \abs{b_e}^2_2 + \abs{W_d}^2_2 +  \abs{b_d}^2_2)$ to avoid over-fitting. We adopt the diffusion operator  $\Phi = \tilde{D}^{-\frac{1}{2}}(\alpha I + (1-\alpha) A)\tilde{D}^{-\frac{1}{2}}$ where $A$ is the adjacency matrix, $D$ is the degree matrix and $\tilde{D} = \alpha I + (1-\alpha) D$ keeps $\Phi$ normalized. The parameter $\alpha \in [0,1]$ tunes nodes' adherence to their previous information or their neighbors' collective opinion. We use it as a hyper-parameter here for simplicity and we can make it as a learnable parameter later.
The differential equation system $\frac{d X}{d t} = \Phi X$ 
follows the dynamics of averaging the neighborhood opinion as $ \frac{d \overrightarrow{x_i(t)}}{d t} = \frac{\alpha}{(1-\alpha)d_i + \alpha}\overrightarrow{x_i(t)} + \sum_j^{n}A_{i,j}\frac{1-\alpha}{\sqrt{(1-\alpha)d_i + \alpha}\sqrt{(1-\alpha)d_j + \alpha}}\overrightarrow{x_j(t)}$
for node $i$. When $\alpha=0$, $\Phi$ averages the neighbors as normalized random walk, when $\alpha=1$, $\Phi$ captures exponential dynamics without network effects, and  when $\alpha=0.5$, $\Phi$ averages both neighbors and itself as  in \cite{KipfW17}. 

\subsection{Experiments}
\mytag{Datasets and Baselines.}
\label{sec:label_data}
\begin{table}
\vspace{-1em}
  \caption{Statistics for three real-world citation network datasets. N, E, D, C represent number of nodes, edges, features, classes respectively.}
  \label{tab:data}
  \centering
  \scriptsize
  \begin{tabular}{llllll}
    \toprule
    Dataset     & N & E & D  & C & Train/Valid/Test\\
    \midrule
    Cora &  $2,708$ & $5,429$  & $1,433$ & $7$  &  $140$/$500$/$1,000$ \\
    Citeseer & $3,327$ & $4,732$ & $3,703$ & $6$ & $120$/$500$/$1,000$\\
    Pubmed &  $19,717$ &  $44,338$ & $500$ & $3$  & $60$/$500$/$1,000$\\
    \bottomrule
  \end{tabular}
  \vspace{-1.5em}
\end{table}
We use standard benchmark datasets, i.e., citation network Cora, Citeseer and Pubmed, and follow the same fixed split scheme for train, validation, and test
as in \cite{YangCS16,KipfW17,thekumparampil2018attention}. We summarize the datasets in 
Table~\ref{tab:data}.
We compare our \model model with 
graph convolution network (GCN) \cite{KipfW17}, attention-based graph neural network (AGNN) \cite{thekumparampil2018attention}, and graph attention networks (GAT) \cite{velivckovic2017graph} with sophisticated attention parameters. 

\mytag{Experimental setup.}
For the consistency of comparison with prior work, we follow the same experimental setup as \cite{KipfW17,velivckovic2017graph,thekumparampil2018attention}. We train our model based on the training datasets and get the accuracy of classification results from the test datasets with $1,000$ labels as summarized in Table~\ref{tab:data}.  Following hyper-parameter settings apply to all the datasets. 
We set $16$ evenly spaced time ticks in $[0, T]$ and solve the initial value problem of integrating the differential equation systems numerically by DOPRI5 \cite{dormand1996numerical}.
We train our model for a maximum of 100 epochs using Adam \cite{KingmaB14} with learning rate $0.01$ and $\ell_2$-norm regularization $0.024$.
We grid search the best terminal time  $T \in [0.5,1.5]$ and the $\alpha \in [0, 1]$. We use $256$ hidden dimension. We report the mean and standard deviation of results for $100$ runs in Table~\ref{tab:results}. It's worthwhile to emphasize that in our model there is no running control parameters (i.e. linear connection layers in GNNs), no dropout (e.g., dropout rate $0.5$ in GCN and $0.6$ in GAT), no early stop, and no concept of layer/network depth (e.g., 2 layers in GCN and GAT). 

\mytag{Results.}
We summarize the results in Table~\ref{tab:results}. We find our \model outperforms many state-of-the-art GNN models. Results for the baselines are taken from \cite{KipfW17,velivckovic2017graph,thekumparampil2018attention,Felix19}.  We report the mean and standard deviation of our results for $100$ runs. We get our reported results in Table~\ref{tab:results} when terminal time $T= 1.2$ , $\alpha= 0 $ for the Cora dataset,  $T=1.0$, $\alpha=0.8$ for the  Citeseer dataset,  and $T=1.1$, $\alpha=0.4$ for the Pubmed dataset. We find best accuracy is achieved at a real-number time/depth.
\vspace{-.5em}
\begin{table}[!h]
\vspace{-0.5em}
  \caption{Test mean accuracy with standard deviation in percentage ($\%$) over $100$ runs. Our \model model gives very competitive results compared with many GNN models. }
  \label{tab:results}
  \centering
\scriptsize
\begin{tabular}{llll}
    \toprule
    Model     & Cora     & Citeseer  & Pubmed \\
    \midrule
    GCN & $81.5$  & $70.3$  &  $79.0$ \\
    AGNN & $83.1\pm0.1$ & $71.7\pm0.1$ & $\mathbf{79.9\pm0.1}$ \\
    GAT & $83.0\pm0.7$ & $72.5\pm0.7$  &  $79.0\pm0.3$ \\
    \bottomrule
    \textbf{\model} &   $\mathbf{83.3\pm0.6}$ & $\mathbf{73.1\pm0.6}$ &  $\mathbf{79.8 \pm 0.4}$\\
    \bottomrule
  \end{tabular}
  \vspace{-1.em}
\end{table}

By capturing the continuous-time network dynamics 
to diffuse features and given labels on graphs, our \model gives better classification accuracy at terminal time $T \in \mathbb{R^+}$. Figure~\ref{fig:gnn_time} plots the mean accuracy  with error bars over terminal time $T$ in the abovementioned $\alpha$ settings (we further plot the accuracy over terminal time $T$ and $\alpha$ in the insets and Appendix~\ref{appendix:accuracy}). We find for all the three datasets their accuracy curves follow rise and fall patterns around the best terminal time stamps which are real number. Indeed, when the terminal time $T$ is too small or too large, the accuracy degenerates because the features of nodes are in under-diffusion or over-diffusion states. The prior GNNs can only have an discrete number of layers which can not capture the continuous-time network dynamics accurately.  In contrast,  our \model captures continuous-time dynamics on graphs in a more fine-grained manner.   

 \section{Conclusion}

 We propose to combine differential equation systems and graph neural networks to learn continuous-time dynamics on complex networks. 
 Our \model gives 
 the  meaning of  physical time and the continuous-time network dynamics to the 
depth and hidden outputs of GNNs respectively, predicts continuous-time dynamics on complex network and regularly-sampled structured sequence  accurately, and outperforms many GNN models in the node semi-supervised classification task (a one-snapshot case). Our model potentially serves as a unified framework to capture the structure and dynamics  of complex systems in a data-driven manner. For future work, we try to apply our model to other applications including molecular dynamics and urban traffics. Codes and datasets are open-sourced  at \url{https://github.com/calvin-zcx/ndcn}.
%
{\small
\vspace{-0.05in}
\section*{Acknowledgement}
\vspace{-0.05in}
This work is supported by NSF 1716432, 1750326, ONR N00014-18-1-2585, Amazon Web Service (AWS) Machine Learning for Research Award and Google Faculty Research Award.
}

\bibliographystyle{ACM-Reference-Format}
\bibliography{acmart}

\section*{Appendix:}
\appendix

\hide{
\section{Reproducibility}
\label{appendix:code}
To ensure the reproducibility, we open-sourced our datasets and Pytorch implementation  empowered by GPU and sparse matrix at:
  \url{https://drive.google.com/open?id=19x7uas9G5w0gU8bHDoohJmDrVOt9wl8W}.
}

\section{Animations of the real-world dynamics on different networks}\label{appendix:animation}
Please view the animations of the three real-world dynamics on five different networks learned by different models at:
  \url{https://drive.google.com/open?id=1KBl-6Oh7BRxcQNQrPeHuKPPI6lndDa5Y}.


\hide{
\subsection{Underlying Networks}
We generate various networks by as follows, and we visualize their adjacency matrix after re-ordering their nodes by the community detection method by Newman \cite{newman2010networks}.
\bit
\item Grid network: 
\begin{figure}[!htb]
\centering
\includegraphics[width=0.5\textwidth]{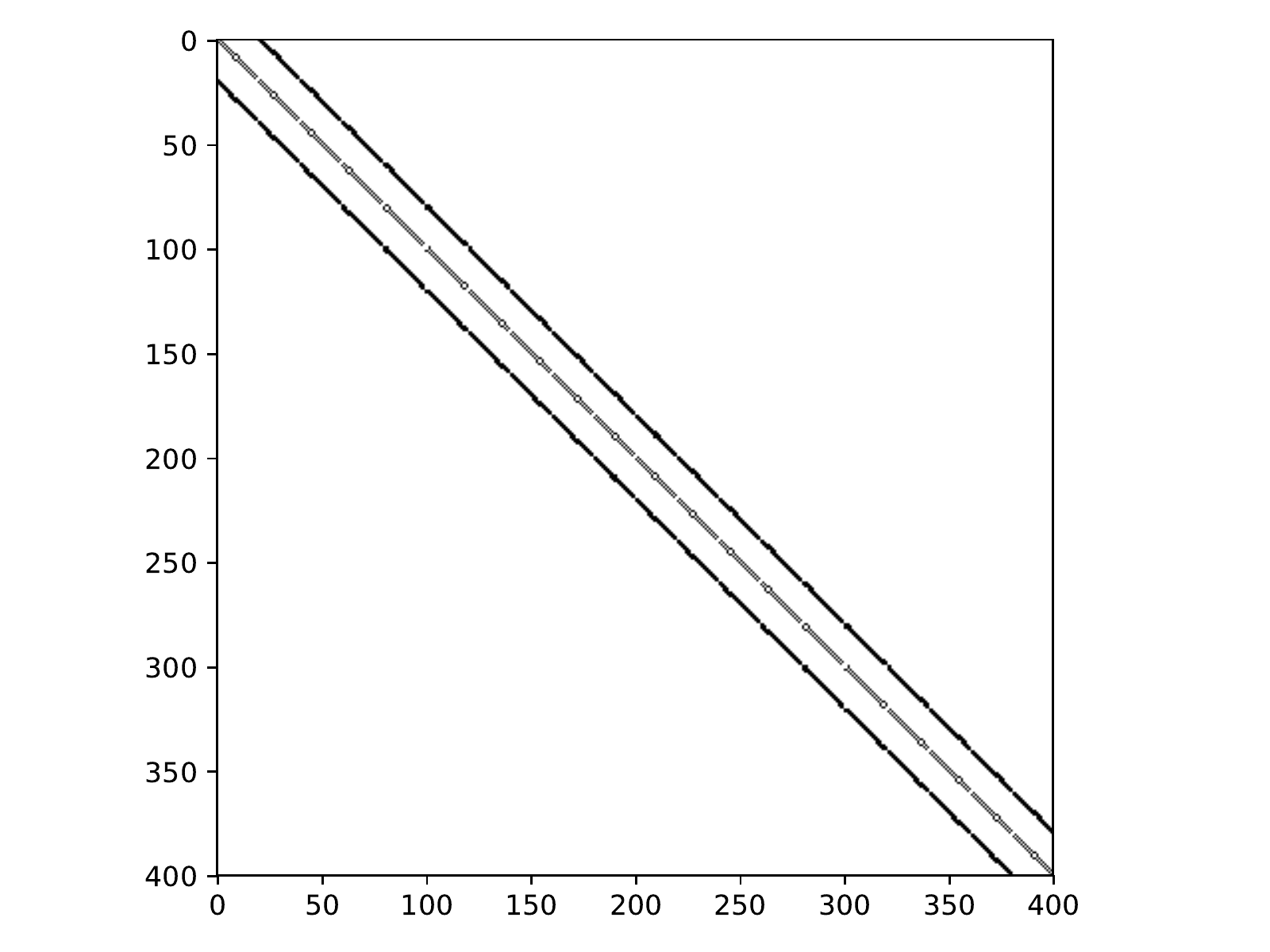}
\caption{ Adjacency matrix of grid network taking on a circulant matrix. } 
\label{fig:appendix:net_grid}
\end{figure}

\hide{
\begin{lstlisting}
    def grid_8_neighbor_graph(N):
        """
        Build discrete grid graph, each node has 8 neighbors
        :param n:  sqrt of the number of nodes
        :return:  A, the adjacency matrix
        """
        N = int(N)
        n = int(N ** 2)
        dx = [-1, 0, 1, -1, 1, -1, 0, 1]
        dy = [-1, -1, -1, 0, 0, 1, 1, 1]
        A = torch.zeros(n, n)
        for x in range(N):
            for y in range(N):
                index = x * N + y
                for i in range(len(dx)):
                    newx = x + dx[i]
                    newy = y + dy[i]
                    if N > newx >= 0 and N > newy >= 0:
                        index2 = newx * N + newy
                        A[index, index2] = 1
        return A.float()
    n = 400
    N = int(np.ceil(np.sqrt(n))) 
    A = grid_8_neighbor_graph(N)
\end{lstlisting}
}

\item Random network:
\begin{figure}[!htb]
\centering
\includegraphics[width=0.5\textwidth]{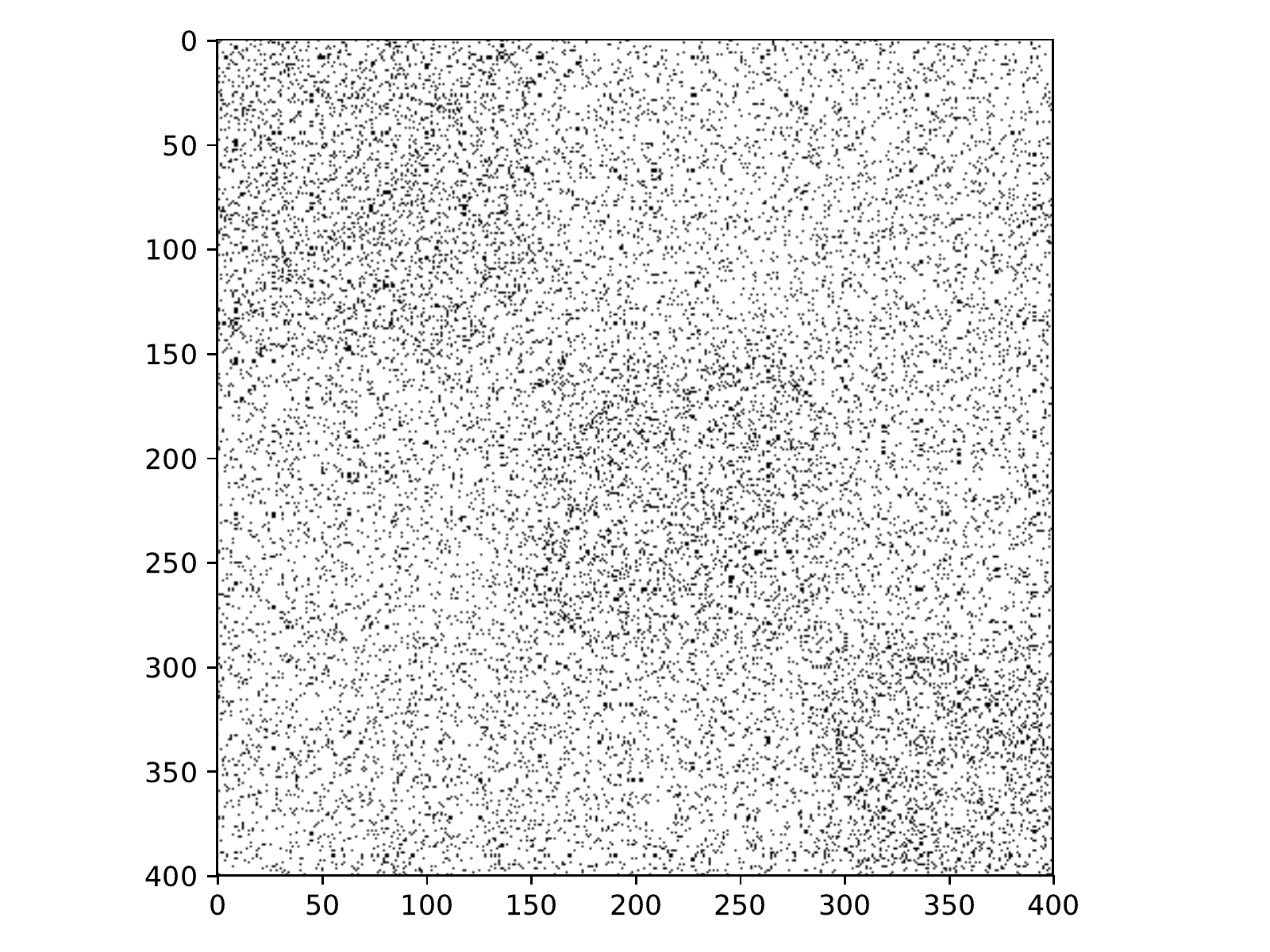}
\caption{ Adjacency matrix of random network. }
\label{fig:appendix:net_random}
\end{figure}
\begin{lstlisting}
import networkx as nx
n = 400
G = nx.erdos_renyi_graph
(n, 0.1, seed=seed)
\end{lstlisting}

\item Power-law network:
\begin{figure}[h]
\centering
\includegraphics[width=0.5\textwidth]{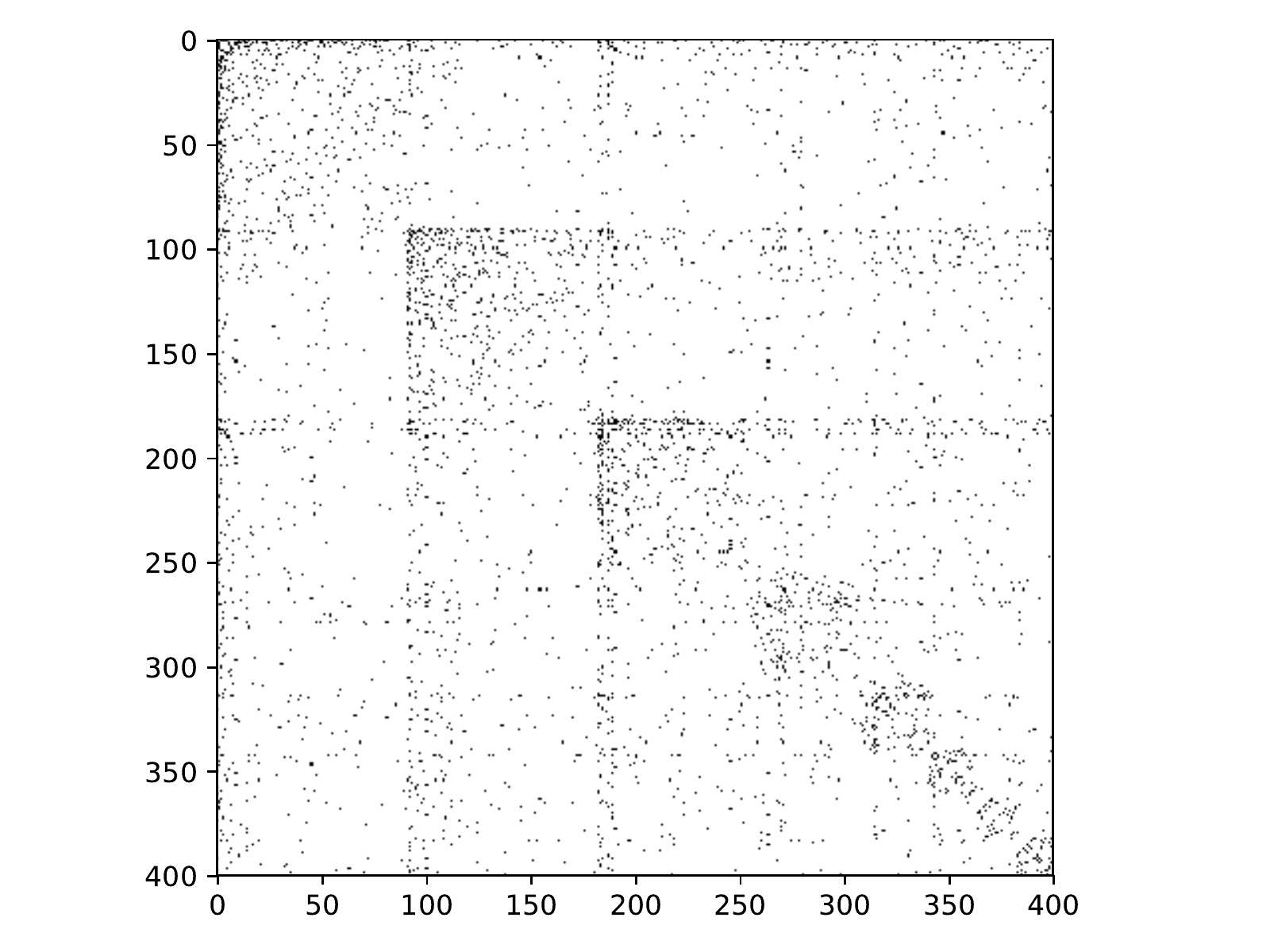}
\caption{ Adjacency matrix of power-law network. }
\label{fig:appendix:net_power_law}
\end{figure}
\begin{lstlisting}
n = 400
G = nx.barabasi_albert_graph
(n, 5, seed=seed)
\end{lstlisting}

\item Small-world network:
\begin{figure}[!htb]
\centering
\includegraphics[width=0.5\textwidth]{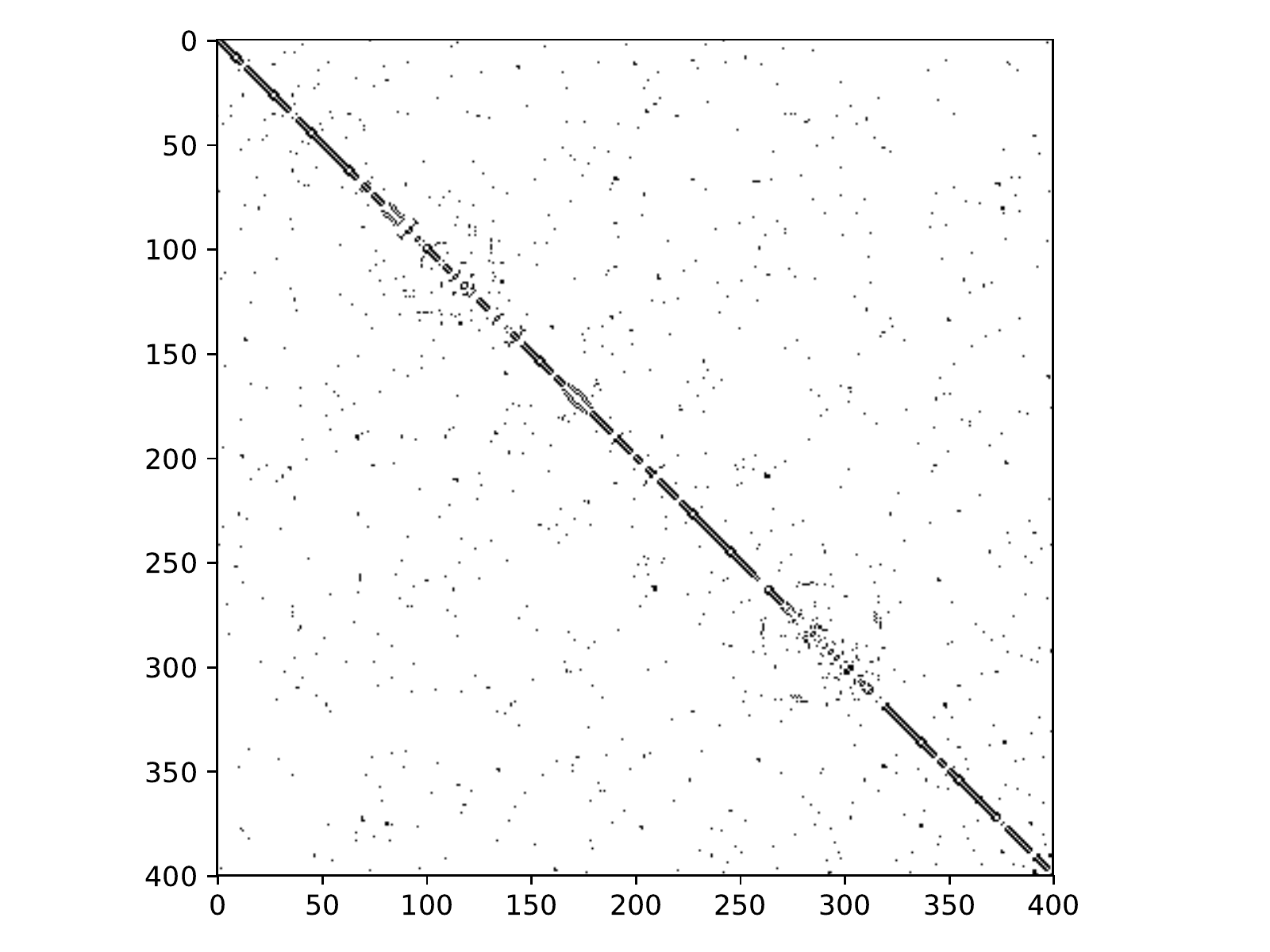}
\caption{ Adjacency matrix of small-world network. }
\label{fig:appendix:net_small_world}
\end{figure}
\begin{lstlisting}
n = 400
G = nx.newman_watts_strogatz_graph
(400, 5, 0.5, seed=seed)
\end{lstlisting}
    
\item Community network:
\begin{figure}[!htb]
\centering
\includegraphics[width=0.5\textwidth]{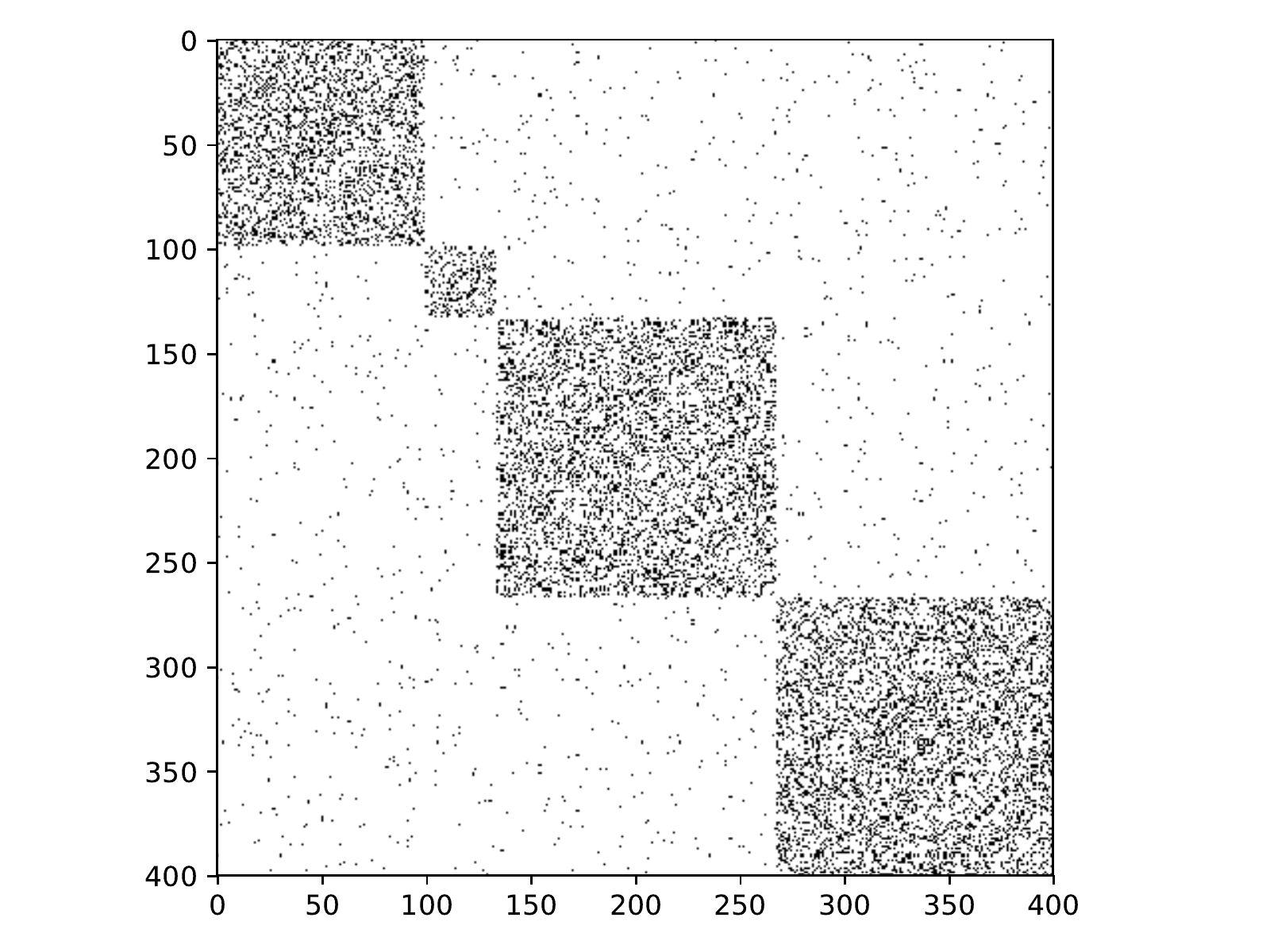}
\caption{ Adjacency matrix of community network. }
\label{fig:appendix:net_community}
\end{figure}
\begin{lstlisting}
n1 = int(n/3)
n2 = int(n/3)
n3 = int(n/4)
n4 = n - n1 - n2 -n3
G = nx.random_partition_graph
([n1, n2, n3, n4], .25, .01, seed=seed)
    \end{lstlisting}

\eit

\subsection{Initial Values of Network Dynamics}
 We set the initial value $X(0)$ the same
for all the experimental settings
and thus different dynamics are only due to their different dynamic rules and underlying networks modelled by $\Dot{X}= f(X, G, W, t)$ as shown in Fig.~\ref{fig:heat},\ref{fig:mutualistic} and \ref{fig:gene}.  Please see above animations to check out different network dynamics.
\small{
\begin{lstlisting}
n = 400
N = int(np.ceil(np.sqrt(n))) 
x0 = torch.zeros(N, N)
x0[int(0.05*N):int(0.25*N), 
    int(0.05*N):int(0.25*N)] = 25  
# x0[1:5, 1:5] = 25  for N = 20 or n= 400 case
x0[int(0.45*N):int(0.75*N), 
    int(0.45*N):int(0.75*N)] = 20  
# x0[9:15, 9:15] = 20 for N = 20 or n= 400 case
x0[int(0.05*N):int(0.25*N), 
    int(0.35*N):int(0.65*N)] = 17  
# x0[1:5, 7:13] = 17 for N = 20 or n= 400 case
\end{lstlisting}
}

\subsection{Network Dynamics}
We adopt the following three real-world dynamics  from different disciplines. Please see above animations to check out the visualization of different network dynamics. The differential equation systems are shown as follows:
\bit
    \item The heat diffusion dynamics  governed by Newton's law of cooling \cite{luikov2012analytical},
    {\small
    \begin{equation}
        \begin{aligned} 
              \frac{d \overrightarrow{x_i(t)}}{d t} =  -k_{i,j} \sum\nolimits_{j=1}^{n} A_{i,j} (\overrightarrow{x_i} - \overrightarrow{x_j})
        \end{aligned}
    \end{equation}}
    states that the rate of heat change of node $i$ is proportional to the difference in the temperatures between $i$ and its neighbors with heat capacity matrix $A$.  We use $k = 1$ here.


    \item The mutualistic interaction dynamics among species in ecology, governed by equation 
    {\small
    \begin{equation}
        \begin{aligned} 
              \frac{d \overrightarrow{x_i(t)}}{d t} = b_i + \overrightarrow{x_i} (1 - \frac{\overrightarrow{x_i}}{k_i})(\frac{\overrightarrow{x_i}}{c_i} -1) + \sum_{j=1}^{n} A_{i,j} \frac{\overrightarrow{x_i} \overrightarrow{x_j}}{d_i + e_i \overrightarrow{x_i} + h_j \overrightarrow{x_j}}.
        \end{aligned}
    \end{equation}}
    The mutualistic differential equation systems \cite{gao2016universal} capture the abundance $\vec{x_i(t)}$ of species $i$, consisting of incoming migration term $b_i$, logistic growth with population capacity $k_i$ \cite{zang2018power} and Allee effect \cite{allee1949principles} with cold-start threshold $c_i$, and mutualistic interaction term with interaction network $A$.  We use $b=0.1$, $k=5.0$, $c=1.0$, $d=5.0$, $e=0.9$, $h=0.1$ here.
    
    \item The gene regulatory dynamics governed by Michaelis-Menten equation 
    {\small
    \begin{equation}
        \begin{aligned} 
              \frac{d \overrightarrow{x_i(t)}}{d t} =  - b_i \vec{x_i}^{f} + \sum\nolimits_{j=1}^{n} A_{i,j}\frac{\overrightarrow{x_j}^{h}}{\overrightarrow{x_j}^{h} + 1}
        \end{aligned}
    \end{equation}}
    where the first term models degradation when $f=1$ or dimerization when $f=2$, and the second term captures genetic
activation tuned by the  Hill coefficient $h$  \cite{gao2016universal}. We adopt $b=1.0$, $f=1.0$, $h=2.0$ here.
\eit

\subsection{Terminal Time:}
We use $T = 5$ for mutualistic dynamics and gene regulatory dynamics over different networks, and {$T=5, 0.1, 0.75, 2, 0.2$} for heat dynamics on the grid, random graph, power-law network, small-world network, and community network respectively due to their different time scale of network dynamics. Please see above animations to check out different network dynamics.

\subsection{Visualizations of network dynamics}

Please see above animations to check out the visualization of different network dynamics.
We generate networks by aforementioned network models with $n=400$ nodes. The nodes are re-ordered according to community detection method by Newman \cite{newman2010networks}. We visualize their adjacency matrices in Fig.~\ref{appendix:fig:heat},\ref{appendix:fig:mutualistic} and \ref{appendix:fig:gene}. We layout these networks in a grid and thus nodes' states $X(t)$ are visualized as functions on the grid. Specifically, the nodes are re-ordered according to community detection method by Newman \cite{newman2010networks} and each node has a unique label from $1$ to $n$. We layout these nodes on a 2-dimensional $\sqrt{n} \times \sqrt{n}$ grid and each grid point $(r,c) \in \mathbb{N}^2$ represents the $i^{th}$ node where $i=r\sqrt{n}+c+1$. Thus, nodes' states $X(t) \in \mathbb{R}^{n\times d}$ when $d=1$ can be visualized as a scalar field function $X: \mathbb{N}^2 \to \mathbb{R}$ over the grid.

\begin{figure*}[h]
\centering
\includegraphics[width=.9\textwidth]{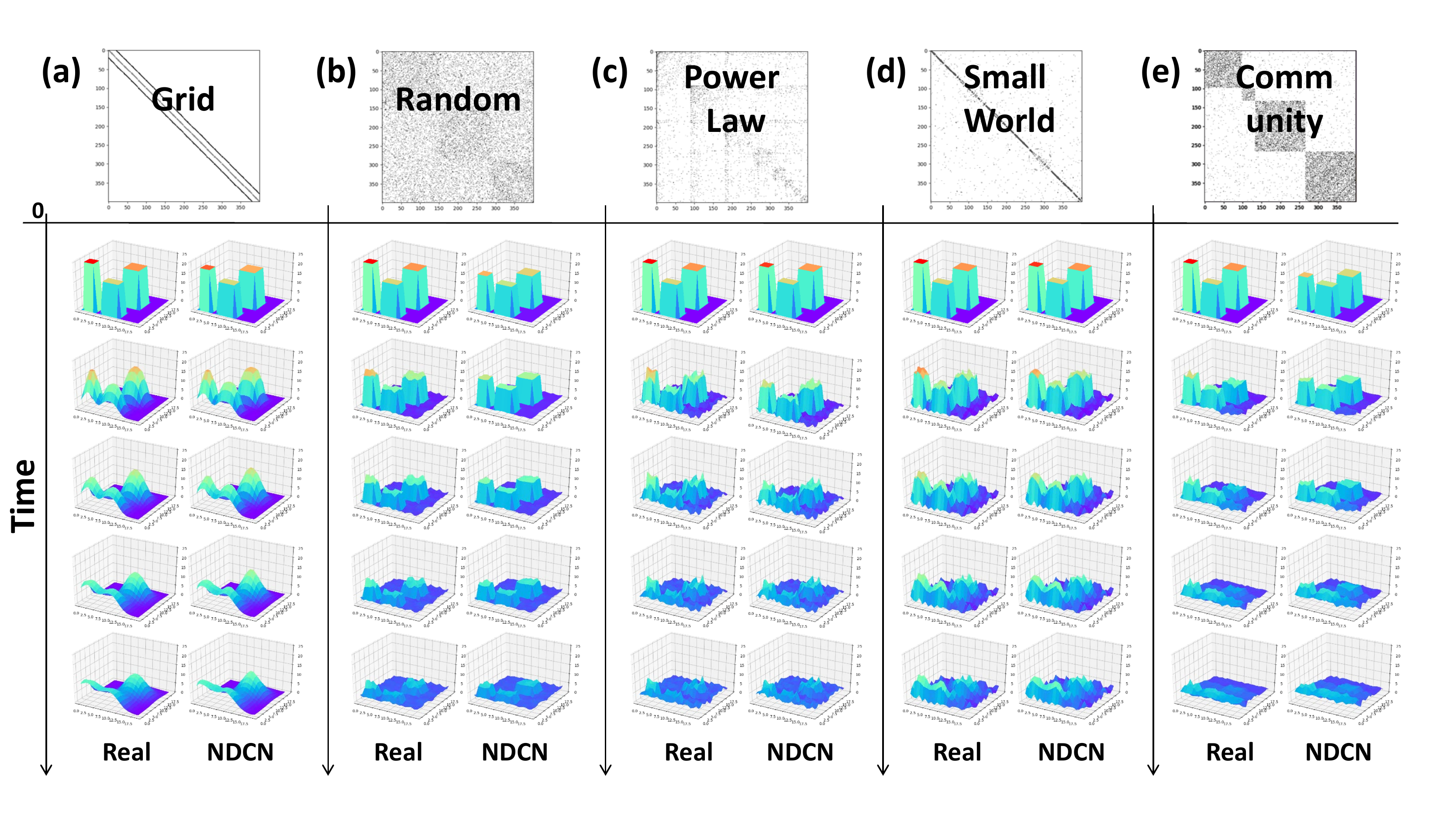}
\caption{ \emph{Heat diffusion on different networks.} 
Each of the five vertical panels represents the dynamics on one network over physical time. For each network dynamics, we illustrate the sampled ground truth dynamics (left) and the dynamics generated by our \model (right) from top to bottom following the direction of time.
}
\label{appendix:fig:heat}
\end{figure*}

\begin{figure*}[h]
\centering
\includegraphics[width=.9\textwidth]{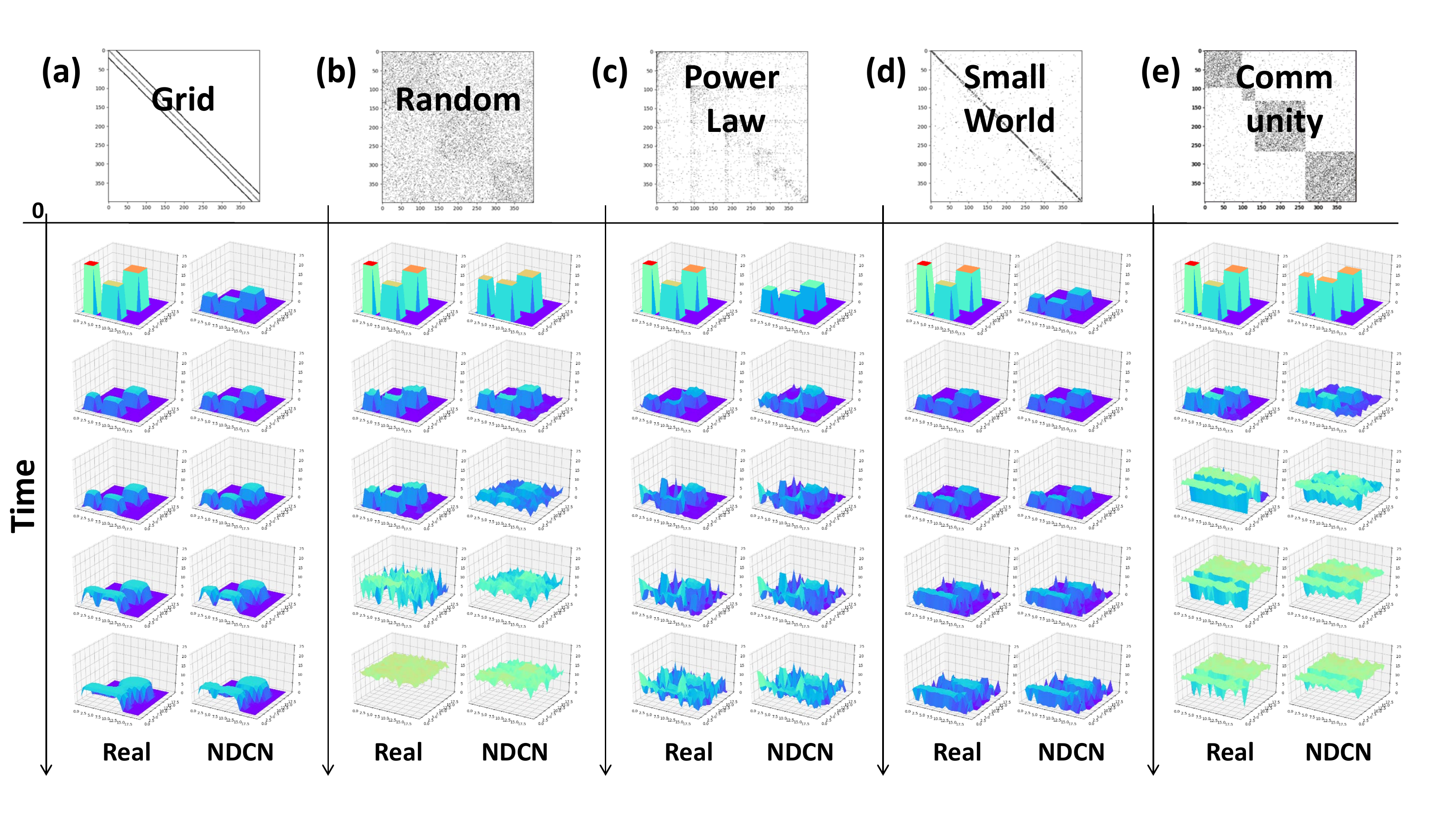}
\caption{ \emph{Biological mutualistic interaction on different networks.} 
}\label{appendix:fig:mutualistic}
\end{figure*}

\begin{figure*}[h]
\centering
\includegraphics[width=.9\textwidth]{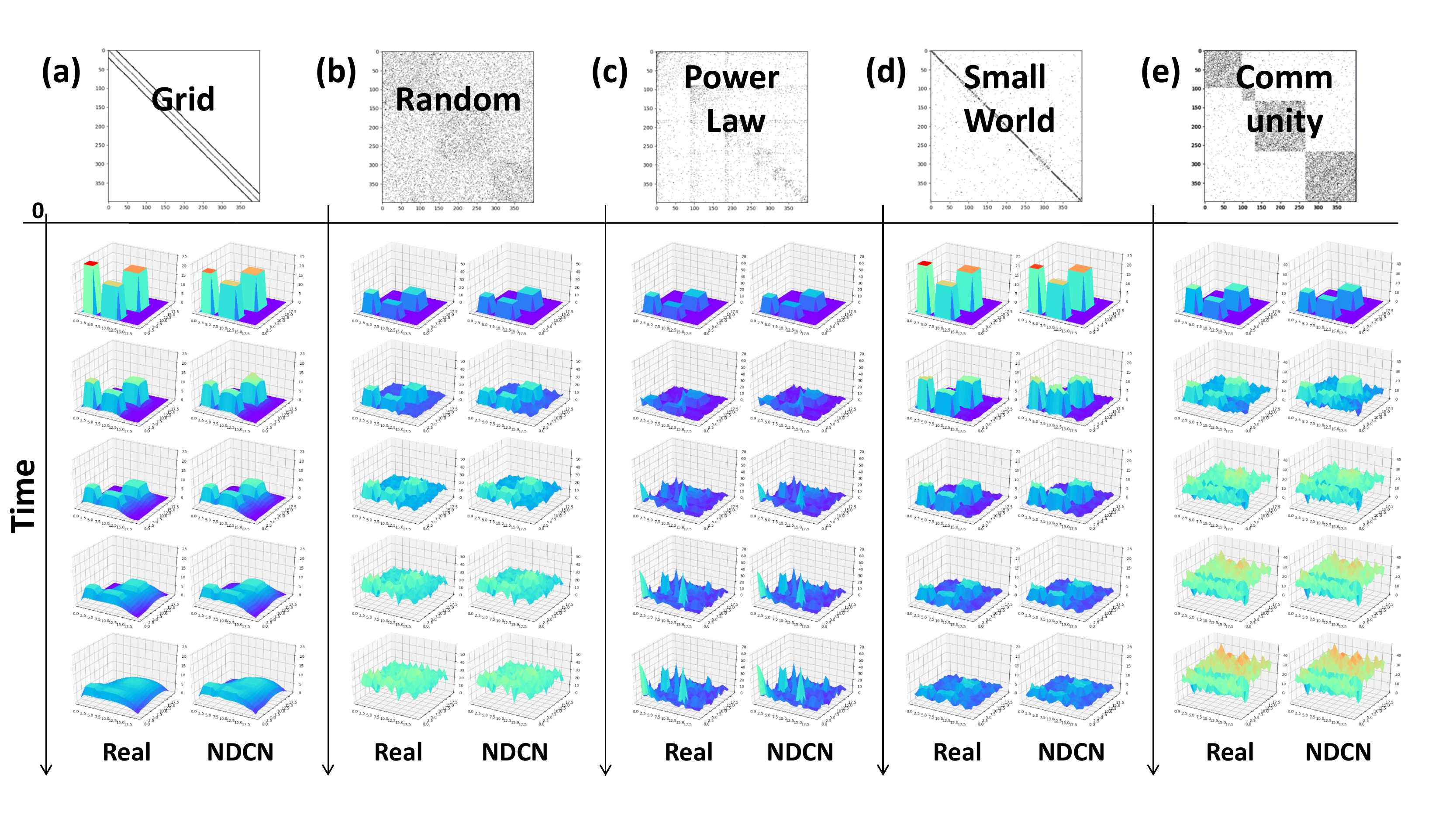}
\caption{ \emph{Gene regulation dynamics on different networks.} 
}
\label{appendix:fig:gene}
\end{figure*}

\hide{
    \begin{lstlisting}
    def generate_node_mapping(G, type=None):
        """
        :param G:
        :param type:
        :return:
        """
        if type == 'degree':
            s = sorted(G.degree, key=lambda x: x[1], reverse=True)
            new_map = {s[i][0]: i for i in range(len(s))}
        elif type == 'community':
            cs = list(community.greedy_modularity_communities(G))
            l = []
            for c in cs:
                l += list(c)
            new_map = {l[i]:i for i in range(len(l))}
        else:
            new_map = None
    
        return new_map


    def networkx_reorder_nodes(G, type=None):
        """
        :param G:  networkX only adjacency matrix without attrs
        :param nodes_map:  nodes mapping dictionary
        :return:
        """
        nodes_map = generate_node_mapping(G, type)
        if nodes_map is None:
            return G
        C = nx.to_scipy_sparse_matrix(G, format='coo')
        new_row = np.array([nodes_map[x] for x in C.row], dtype=np.int32)
        new_col = np.array([nodes_map[x] for x in C.col], dtype=np.int32)
        new_C = sp.coo_matrix((C.data, (new_row, new_col)), shape=C.shape)
        new_G = nx.from_scipy_sparse_matrix(new_C)
        return new_G
        
    G = networkx_reorder_nodes(G, args.layout)
    A = torch.FloatTensor(nx.to_numpy_array(G))
    
    def visualize_graph_matrix(G, title, dir=r'figure/network'):
        A = nx.to_numpy_array(G)
        fig = plt.figure()  # figsize=(12, 4), facecolor='white'
        fig.tight_layout()
        plt.imshow(A, cmap='Greys')  # ''YlGn')
        # plt.pcolormesh(A)
        plt.show()
    
    def visualize_network_dynamics(N, x0, xt, figname, title ='Dynamics in Complex Network', dir='png_learn_dynamics', zmin=None, zmax=None):
        """
        :param N:   N**2 is the number of nodes, N is the pixel of grid
        :param x0:  initial condition
        :param xt:  states at time t to plot
        :param figname:  figname , numbered
        :param title: title in figure
        :param dir: dir to save
        :param zmin: ax.set_zlim(zmin, zmax)
        :param zmax: ax.set_zlim(zmin, zmax)
        :return:
        """
        if zmin is None:
            zmin = x0.min()
        if zmax is None:
            zmax = x0.max()
        fig = plt.figure()  # figsize=(12, 4), facecolor='white'
        fig.tight_layout()
        x0 = x0.detach()
        xt = xt.detach()
        ax = fig.gca(projection='3d')
        ax.cla()
        X = np.arange(0, N)
        Y = np.arange(0, N)
        X, Y = np.meshgrid(X, Y) 
        surf = ax.plot_surface(X, Y, xt.numpy().reshape((N, N)), cmap='rainbow',
                               linewidth=0, antialiased=False, vmin=zmin, vmax=zmax)
        ax.set_zlim(zmin, zmax)
        fig.savefig(dir+'/'+figname+".png", transparent=True)
        fig.savefig(dir+'/'+figname + ".pdf", transparent=True)
        # plt.draw()
        plt.pause(0.001)
        plt.close(fig)
\end{lstlisting}
}
}

\section{Model configurations}
\label{appendix:conf}
Model configurations of learning network dynamics in both continuous-time and regularly-sampled settings.
We train our \model model by Adam \cite{KingmaB14}. We choose $20$ as the hidden dimension of $X_h \in \mathbb{R}^{n\times 20} $.  
We train our model for a maximum of 2000 epochs using Adam \cite{KingmaB14} with learning rate $0.01$. We summarize our $\ell_2$ regularization parameter as in Table~\ref{appendix:table:con_ext_para} and Table~\ref{appendix:table:con_int_para} for Section~\ref{sec:continuous} learning continuous-time network dynamics.
We summarize our $\ell_2$ regularization parameter as in Table~\ref{appendix:table:irr_ext_para} for Section~\ref{sec:regular} learning regularly-sampled dynamics.
 
 \begin{table}[h]
\vspace{-1em}
\scriptsize
\centering
\caption{$\ell_2$ regularization parameter configurations in continuous-time extrapolation prediction
 }
\label{appendix:table:con_ext_para}
\begin{tabular}{l l l l l l l}
\hline
  & & Grid & Random & Power Law & Small World & Community \\ \hline
\multirow{4}{*}{\makecell{Heat \\ Diffusion}} 
 & No-Encode  &  1e-3 & 1e-6 & 1e-3 & 1e-3 & 1e-5\\ 
 & No-Graph  &  1e-3 & 1e-6 & 1e-3 & 1e-3 & 1e-5\\ 
 & No-Control  &  1e-3 & 1e-6 & 1e-3 & 1e-3 & 1e-5\\ 
 & \textbf{\model}  &  1e-3 & 1e-6 & 1e-3 & 1e-3 & 1e-5\\  \hline 
\multirow{4}{*}{\makecell{Mutualistic \\ Interaction}} 
 & No-Encode  &  1e-2 & 1e-4 & 1e-4 & 1e-4 & 1e-4\\ 
 & No-Graph  &  1e-2 & 1e-4 & 1e-4 & 1e-4 & 1e-4\\
 & No-Control  &  1e-2 & 1e-4 & 1e-4 & 1e-4 & 1e-4\\
 & \textbf{\model}  &  1e-2 & 1e-4 & 1e-4 & 1e-4 & 1e-4\\ \hline 
\multirow{4}{*}{\makecell{Gene \\ Regulation}} 
 & No-Embed  &  1e-4 & 1e-4 & 1e-4 & 1e-4 & 1e-4\\
 & No-Graph  &  1e-4 & 1e-4 & 1e-4 & 1e-4 & 1e-4\\
 & No-Control  &  1e-4 & 1e-4 & 1e-4 & 1e-4 & 1e-4\\
 & \textbf{\model}  &  1e-4 & 1e-4 & 1e-4 & 1e-4 & 1e-4\\ \hline
\end{tabular}
\vspace{-1em}
\end{table}

\begin{table}[h]
\vspace{-1em}
\scriptsize
\centering
\caption{$\ell_2$ regularization parameter configurations in continuous-time interpolation prediction
 }
\label{appendix:table:con_int_para}
\begin{tabular}{l l l l l l l}
\hline
  & & Grid & Random & Power Law & Small World & Community \\ \hline
\multirow{4}{*}{\makecell{Heat \\ Diffusion}} 
 & No-Encode  &  1e-3 & 1e-6 & 1e-3 & 1e-3 & 1e-5\\ 
 & No-Graph  &  1e-3 & 1e-6 & 1e-3 & 1e-3 & 1e-5\\ 
 & No-Control  &  1e-3 & 1e-6 & 1e-3 & 1e-3 & 1e-5\\ 
 & \textbf{\model}  &  1e-3 & 1e-6 & 1e-3 & 1e-3 & 1e-5\\  \hline 
\multirow{4}{*}{\makecell{Mutualistic \\ Interaction}} 
 & No-Encode  &  1e-2 & 1e-4 & 1e-4 & 1e-4 & 1e-4\\ 
 & No-Graph  &  1e-2 & 1e-4 & 1e-4 & 1e-4 & 1e-4\\
 & No-Control  &  1e-2 & 1e-4 & 1e-4 & 1e-4 & 1e-4\\
 & \textbf{\model}  &  1e-2 & 1e-4 & 1e-4 & 1e-4 & 1e-4\\ \hline 
\multirow{4}{*}{\makecell{Gene \\ Regulation}} 
 & No-Embed  &  1e-4 & 1e-4 & 1e-4 & 1e-4 & 1e-4\\
 & No-Graph  &  1e-4 & 1e-4 & 1e-4 & 1e-4 & 1e-4\\
 & No-Control  &  1e-4 & 1e-4 & 1e-4 & 1e-4 & 1e-4\\
 & \textbf{\model}  &  1e-4 & 1e-4 & 1e-4 & 1e-4 & 1e-4\\ \hline
\vspace{-1em}
\end{tabular}
\end{table}

\begin{table}[h]
\vspace{-1em}
\tiny
\centering
\caption{$\ell_2$ regularization parameter configurations in regularly-sampled extrapolation prediction
 }
\label{appendix:table:irr_ext_para}
\begin{tabular}{l l l l l l l}
\hline
  & & Grid & Random & Power Law & Small World & Community \\ \hline
\multirow{4}{*}{\makecell{Heat \\ Diffusion}} 
 & LSTM-GNN  &  1e-3 & 1e-3 & 1e-3 & 1e-3 & 1e-3\\ 
 & GRU-GNN  &  1e-3 & 1e-3 & 1e-3 & 1e-3 & 1e-3\\ 
 & RNN-GNN  &  1e-3 & 1e-3 & 1e-3 & 1e-3 & 1e-3\\ 
 & \textbf{\model}  &  1e-3 & 1e-6 & 1e-3 & 1e-3 & 1e-5\\  \hline 
\multirow{4}{*}{\makecell{Mutualistic \\ Interaction}} 
 & LSTM-GNN  &  1e-3 & 1e-3 & 1e-3 & 1e-3 & 1e-3\\ 
 & GRU-GNN  &  1e-3 & 1e-3 & 1e-3 & 1e-3 & 1e-3\\
 & RNN-GNN  &  1e-3 & 1e-3 & 1e-3 & 1e-3 & 1e-3\\
 & \textbf{\model}  &  1e-2 & 1e-3 & 1e-4 & 1e-4 & 1e-4\\ \hline 
\multirow{4}{*}{\makecell{Gene \\ Regulation}} 
 & LSTM-GNN  &  1e-3 & 1e-3 & 1e-3 & 1e-3 & 1e-3\\
 & GRU-GNN  &  1e-3 & 1e-3 & 1e-3 & 1e-3 & 1e-3\\
 & RNN-Control  &  1e-3 & 1e-3 & 1e-3 & 1e-3 & 1e-3\\
 & \textbf{\model}  &  1e-4 & 1e-4 & 1e-4 & 1e-3 & 1e-3\\ \hline
\vspace{-1em}
\end{tabular}
\end{table}

\hide{
\section{Temporal-GNN models}
\label{appendix:rnn}
We use following temporal-GNN models for structured sequence learning:

 \bit
    \item LSTM-GNN: the temporal-GNN with LSTM cell: 
\begin{equation}
\begin{aligned} 
x_t &= ReLU(\Phi*(W_e *X[t]+b_e)) \\
i_t &= \sigma(W_{ii} x_t + b_{ii} + W_{hi} h_{t-1} + b_{hi}) \\
f_t &= \sigma(W_{if} x_t + b_{if} + W_{hf} h_{t-1} + b_{hf}) \\
g_t &= \tanh(W_{ig} x_t + b_{ig} + W_{hg} h_{t-1} + b_{hg}) \\
o_t &= \sigma(W_{io} x_t + b_{io} + W_{ho} h_{t-1} + b_{ho}) \\
c_t &= f_t * c_{t-1} + i_t * g_t \\
h_t &= o_t * \tanh(c_t) \\
\hat{X[t+1]} &= W_d * h_t + b_d\\
\end{aligned}
\end{equation}

    \item GRU-GNN: the temporal-GNN with GRU cell:  
\begin{equation}
\begin{aligned} 
x_t &= ReLU(\Phi*(W_e *X[t]+b_e)) \\
r_t &= \sigma(W_{ir} x_t + b_{ir} + W_{hr} h_{t-1} + b_{hr}) \\
z_t &= \sigma(W_{iz} x_t + b_{iz} + W_{hz} h_{t-1} + b_{hz}) \\
n_t &= \tanh(W_{in} x_t + b_{in} + r * (W_{hn} h_{t-1} + b_{hn})) \\
h_t &= (1 - z_t) * n_t + z_t * h_{t-1}\\
\hat{X[t+1]} &= W_d * h_t + b_d\\
\end{aligned}
\end{equation}
    
    \item RNN-GNN: the temporal-GNN with RNN cell:  
\begin{equation}
\begin{aligned} 
x_t &= ReLU(\Phi*(W_e *X[t]+b_e)) \\   
h_t &= \tanh(w_{ih} x_t + b_{ih}  +  w_{hh} h_{t-1} + b_{hh})\\
\hat{X[t+1]} &= W_d * h_t + b_d\\
\end{aligned}
\end{equation}

\eit
We adopt the diffusion operator  $\Phi = \tilde{D}^{-\frac{1}{2}}(\alpha I + (1-\alpha) A)\tilde{D}^{-\frac{1}{2}}$ where $A$ is the adjacency matrix, $D$ is the degree matrix and $\tilde{D} = \alpha I + (1-\alpha) D$ keeps $\Phi$ normalized.  
The differential equation system $\frac{d X}{d t} = \Phi X$ 
follows the dynamics of averaging the neighborhood opinion as $ \frac{d \overrightarrow{x_i(t)}}{d t} = \frac{\alpha}{(1-\alpha)d_i + \alpha}\overrightarrow{x_i(t)} + \sum_j^{n}A_{i,j}\frac{1-\alpha}{\sqrt{(1-\alpha)d_i + \alpha}\sqrt{(1-\alpha)d_j + \alpha}}\overrightarrow{x_j(t)}$
for node $i$. When $\alpha=0$, $\Phi$ averages the neighbors as normalized random walk, when $\alpha=1$, $\Phi$ captures exponential dynamics without network effects.
Here we adopt $\alpha=0.5$, namely $\Phi$ averages both neighbors and itself as GCN in \cite{KipfW17}. 
}

\section{Results in absolute error.}
\label{appendix:abs}
We show corresponding $\ell_1$ loss error in Table~\ref{table:fit_dyn_abs},Table~\ref{table:irr_inter_abs} and Table~\ref{table:equal_ex_abs} with respect to the normalized $\ell_1$ loss error in Section~\ref{sec:continuous} learning continuous-time network dynamics and Section~\ref{sec:regular} learning regularly-sampled dynamics.
The same conclusions can be made as in Table~\ref{table:fit_dyn},Table~\ref{table:irr_inter} and Table~\ref{table:equal_ex}.


\begin{table}[h]
\vspace{-.5em}
\centering
\caption{Continuous-time Extrapolation Prediction. Our \model predicts different continuous-time network accurately. Each result is the $\ell_1$ error with standard deviation from $20$ runs for 3 dynamics on 5 networks for each method.
 }
\label{table:fit_dyn_abs}
\tiny
\begin{tabular}{l l l l l l l}
\hline
  & & Grid & Random & Power Law & Small World & Community \\ \hline
\multirow{4}{*}{\makecell{Heat \\ Diffusion}} 
 & No-Encode  &  $1.143\pm 0.280$ & $1.060 \pm 0.195$ & $0.950 \pm 0.199$& $0.948 \pm 0.122$& $1.154 \pm 0.167$\\ 
 & No-Graph  & $1.166 \pm 0.066$ &$0.223\pm 0.049$ &$0.260 \pm 0.020$ & $ 0.410\pm 0.023$& $ 0.926\pm 0.116 $\\
 & No-Control  &  $ 2.803\pm 0.549 $& $1.076\pm 0.153$& $0.962\pm 0.163$& $ 1.176\pm 0.179$&$1.417 \pm 0.140$\\
 & \textbf{\model}  &  $\mathbf{ 0.158\pm 0.047}$ & $\mathbf{ 0.163 \pm 0.060}$ & $\mathbf{ 0.187 \pm 0.020}$ & $\mathbf{ 0.097 \pm 0.016}$ & $\mathbf{0.183 \pm 0.039}$\\ \hline 
\multirow{4}{*}{\makecell{Mutualistic \\ Interaction}} 
 & No-Encode  &  $ 1.755\pm 0.138$ & $ 1.402\pm 0.456$ & $ 2.632 \pm 0.775$ & $1.947 \pm 0.106$ & $2.007 \pm 0.695$\\
 & No-Graph  &  $ 2.174 \pm 0.089$ & $ 1.038 \pm 0.434$ & $1.301 \pm 0.551$ & $1.936 \pm 0.085$ & $ 1.323 \pm 0.204$\\
 & No-Control  &  $5.434 \pm 0.473$ & $1.669 \pm 0.662$ & $9.353 \pm 3.751$ & $4.111 \pm 0.417$ & $2.344 \pm 0.424$\\
 & \textbf{\model}  &  $\mathbf {1.038 \pm 0.181 }$ & $\mathbf{ 0.584 \pm 0.277}$ & $\mathbf{ 0.653 \pm 0.230}$ & $\mathbf{ 0.521 \pm 0.124}$ & $\mathbf{ 0.502 \pm 0.210}$\\ \hline 
\multirow{4}{*}{\makecell{Gene \\ Regulation}} 
 & No-Encode  &  $ 2.164 \pm 0.957$ & $ 6.954 \pm 5.190$ & $3.240 \pm 0.954$ & $1.445 \pm 0.395$ & $8.204 \pm 3.240$\\
 & No-Graph  &  $ 0.907 \pm 0.058$ & $4.872 \pm 0.078$ & $4.206 \pm 0.025$ & $0.875 \pm 0.016$ & $6.112 \pm 0.143$\\
 & No-Control  &  $4.458 \pm 0.978$ & $27.119 \pm 2.608$ & $6.768 \pm 0.741$ & $3.320 \pm 0.982$ & $20.002 \pm 2.160$\\
 & \textbf{\model}  & $\mathbf{  1.089 \pm 0.487}$ & $\mathbf{ 0.715 \pm 0.210}$ & $\mathbf{ 0.342 \pm 0.088}$ & $\mathbf{ 0.243 \pm 0.051}$ & $\mathbf{ 0.782 \pm 0.199}$\\ \hline
\end{tabular}
\end{table}

\begin{table}[h]
\vspace{-1em}
\centering
\caption{ Continuous-time Interpolation Prediction. Our \model predicts different continuous-time network accurately. Each result is the $\ell_1$ error with standard deviation from $20$ runs for 3 dynamics on 5 networks for each method.
 }
\label{table:irr_inter_abs}
\tiny
\begin{tabular}{l l l l l l l}
\hline
  & & Grid & Random & Power Law & Small World & Community \\ \hline
\multirow{4}{*}{\makecell{Heat \\ Diffusion}} 
 & No-Encode  &  $1.222 \pm 0.486$ & $1.020 \pm 0.168$ & $0.982 \pm 0.143$ & $1.066 \pm 0.280$ & $1.336 \pm 0.239$\\ 
 & No-Graph  & $1.600 \pm 0.068$ & $0.361\pm 0.022$ & $0.694 \pm 0.058$ & $ 0.956 \pm 0.079$ & $ 0.954\pm 0.053 $\\
 & No-Control  &  $ 2.169\pm 0.108 $& $1.230\pm 0.266$& $1.280\pm 0.216$& $ 1.544\pm 0.128$ & $1.495 \pm 0.171$\\
 & \textbf{\model}  &  $\mathbf{ 0.121\pm 0.024}$ & $\mathbf{ 0.121 \pm 0.017}$ & $\mathbf{ 0.214 \pm 0.024}$ & $\mathbf{ 0.129 \pm 0.017}$ & $\mathbf{0.165 \pm 0.019}$\\ \hline 
\multirow{4}{*}{\makecell{Mutualistic \\ Interaction}} 
 & No-Encode  &  $ 0.620\pm 0.081$ & $ 2.424 \pm 0.598$ & $ 1.755 \pm 0.560$ & $0.488 \pm 0.077$ & $2.777 \pm 0.773$\\
 & No-Graph  &  $ 0.626 \pm 0.143$ & $ 0.967 \pm 0.269$ & $1.180 \pm 0.171$ & $0.497 \pm 0.101$ & $ 1.578 \pm 0.244$\\
 & No-Control  &  $1.534 \pm 0.158$ & $2.836 \pm 1.022$ & $3.328 \pm 0.314$ & $1.212 \pm 0.116$ & $3.601 \pm 0.940$\\
 & \textbf{\model}  &  $\mathbf { 0.164 \pm 0.031 }$ & $\mathbf{ 0.843 \pm 0.267}$ & $\mathbf{ 0.333 \pm 0.055}$ & $\mathbf{ 0.085 \pm 0.014}$ & $\mathbf{ 0.852 \pm 0.247}$\\ \hline 
\multirow{4}{*}{\makecell{Gene \\ Regulation}} 
 & No-Encode  &  $ 1.753 \pm 0.555$ & $ 4.278 \pm 3.374$ & $2.560 \pm 0.765$ & $1.180 \pm 0.389$ & $5.106 \pm 2.420$\\
 & No-Graph  &  $ 1.140 \pm 0.101$ & $3.768 \pm 0.316$ & $3.137 \pm 0.264$ & $0.672 \pm 0.050$ & $4.639 \pm 0.399$\\
 & No-Control  &  $3.010 \pm 0.228$ & $9.939 \pm 1.185$ & $3.139 \pm 0.313$ & $2.082 \pm 0.293$ & $8.659 \pm 0.952$\\
 & \textbf{\model}  & $\mathbf{  0.262 \pm 0.046}$ & $\mathbf{ 0.455 \pm 0.174}$ & $\mathbf{ 0.222 \pm 0.034}$ & $\mathbf{ 0.180 \pm 0.032}$ & $\mathbf{ 0.562 \pm 0.130}$\\ \hline
\end{tabular}
\end{table}

\begin{table}[h]
\vspace{-1em}
\centering
\caption{ Regularly-sampled Extrapolation Prediction. Our \model predicts different structured sequences accurately. Each result is the   $\ell_1$ error with standard deviation  from $20$ runs for 3 dynamics on 5 networks for each method.
 }
\label{table:equal_ex_abs}
\tiny
\begin{tabular}{l l l l l l l}
\hline
  & & Grid & Random & Power Law & Small World & Community \\ \hline
\multirow{4}{*}{\makecell{Heat \\ Diffusion}} 
 & LSTM-GNN &  $0.489 \pm 0.081$ & $0.824 \pm 0.294$ & $0.475 \pm 0.196$ & $0.442 \pm 0.083$ & $0.517 \pm 0.162$\\ 
 & GRU-GNN  & $0.428 \pm 0.085$ & $0.349\pm 0.090$ & $ 0.337 \pm 0.049$ & $ 0.357 \pm 0.065$ & $ 0.302\pm 0.031 $\\
 & RNN-GNN  &  $ 0.717\pm 0.227 $& $0.957\pm 0.215$ & $0.722\pm 0.247$& $ 0.833\pm 0.145$ & $0.615 \pm 0.000$\\
 & \textbf{\model}  &  $\mathbf{ 0.165\pm 0.027}$ & $\mathbf{ 0.180 \pm 0.063}$ & $\mathbf{ 0.208 \pm 0.015}$ & $\mathbf{ 0.103 \pm 0.014}$ & $\mathbf{0.201 \pm 0.029}$\\ \hline 
\multirow{4}{*}{\makecell{Mutualistic \\ Interaction}} 
 & LSTM-GNN  &  $ 1.966\pm 0.126$ & $ 3.749 \pm 3.749$ & $ 2.380 \pm 0.626$ & $2.044\pm 0.086$ & $3.463 \pm 3.095$\\
 & GRU-GNN  &  $ 1.905 \pm 0.157$ & $ \mathbf{0.162 \pm 0.564}$ & $1.077 \pm 0.071$ & $1.792 \pm 0.165$ & $ \mathbf{0.510 \pm 0.549}$\\
 & RNN-GNN  &  $2.165 \pm 0.004$ & $1.303 \pm 1.747$ & $1.056 \pm 0.034$ & $2.012 \pm 0.065$ & $1.140 \pm 0.887$\\
 & \textbf{\model}  &  $\mathbf { 1.414 \pm 0.060 }$ & $\mathbf{ 0.734 \pm 0.168}$ & $\mathbf{ 0.990 \pm 0.442}$ & $\mathbf{ 0.557 \pm 0.078}$ & $\mathbf{ 0.528 \pm 0.122}$\\ \hline 
\multirow{4}{*}{\makecell{Gene \\ Regulation}} 
 & LSTM-GNN  &  $ 1.883 \pm 0.218$ & $ 26.750 \pm 5.634$ & $3.733 \pm 1.220$ & $0.743 \pm 0.112$ & $16.534 \pm 5.094$\\
 & GRU-GNN  &  $ 1.641 \pm 0.191$ & $20.240 \pm 2.549$ & $3.381 \pm 1.455$ & $0.626 \pm 0.099$ & $14.4 \pm 2.358$\\
 & RNN-GNN  &  $1.906 \pm 0.464$ & $22.46 \pm 2.276$ & $4.036 \pm 1.229$ & $0.795 \pm 0.300$ & $14.496 \pm 1.077$\\
 & \textbf{\model}  & $\mathbf{  1.267 \pm 0.672}$ & $\mathbf{ 0.946 \pm 0.357}$ & $\mathbf{ 0.397 \pm 0.133}$ & $\mathbf{ 0.312 \pm 0.043}$ & $\mathbf{ 0.901 \pm 0.160}$\\ \hline
\end{tabular}
\end{table}

\section{Accuracy over terminal time and $\alpha$}
\label{appendix:accuracy}
By capturing the continuous-time network dynamics, our \model gives better classification accuracy at terminal time $T \in \mathbb{R^+}$. Indeed, when the terminal time is too small or too large, the accuracy degenerates because the features of nodes are in under-diffusion or over-diffusion states. 
We plot the mean accuracy of 100 runs of our \model model over different terminal time $T$ and $\alpha$ as shown in the following heatmap plots. we find for all the three datasets their accuracy curves follow rise and fall pattern around the best terminal time. 
\begin{figure}[h]
\vspace{-0.5em}
\centering
\includegraphics[width=0.35\textwidth]{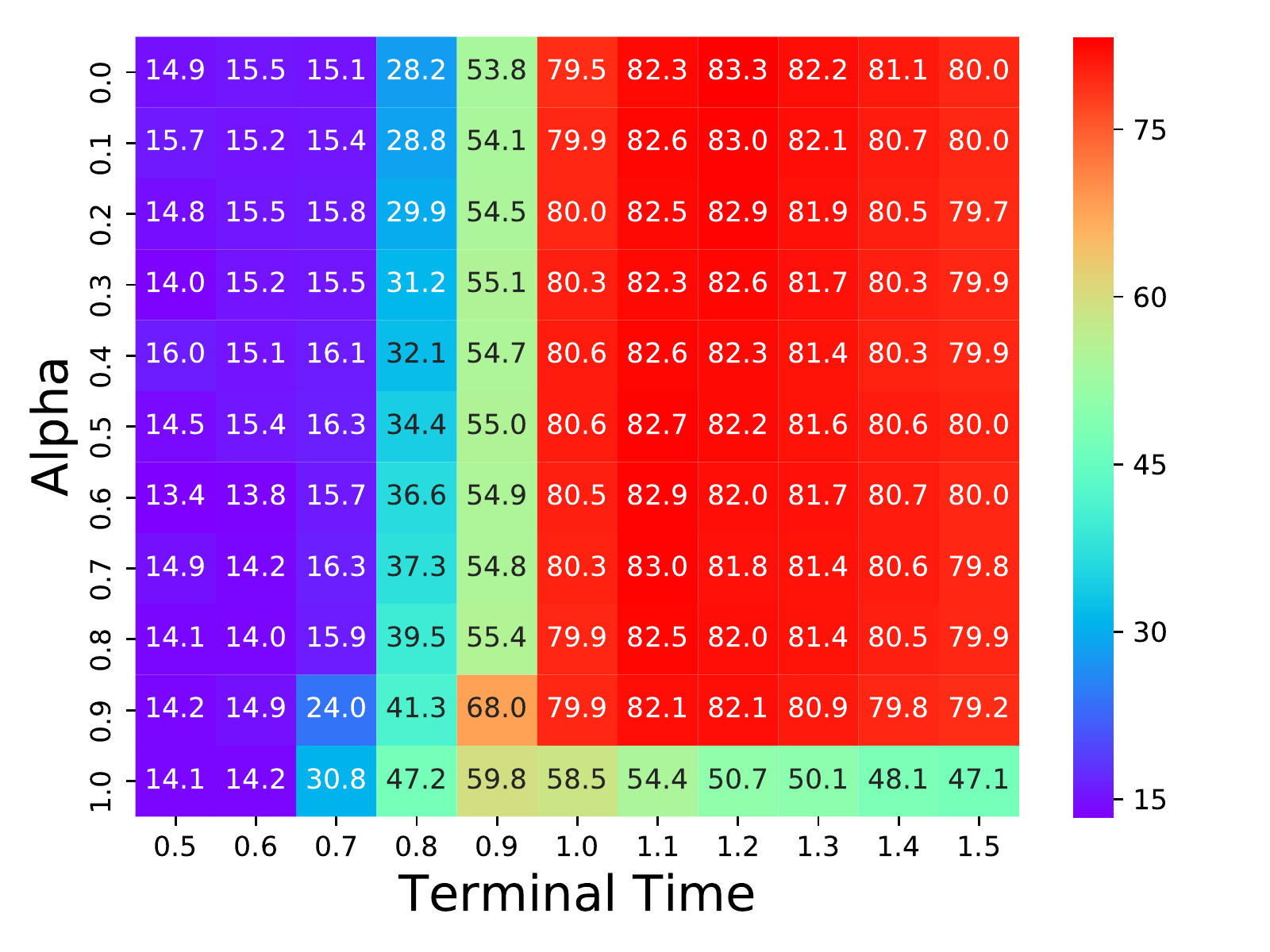}
\caption{  Mean classification accuracy of $100$ runs of our \model model over terminal time and $\alpha$ for the Cora dataset in heatmap plot. }
\label{fig:appendix:cora}
\end{figure}

\begin{figure}[h]
\vspace{-0.5em}
\centering
\includegraphics[width=0.35\textwidth]{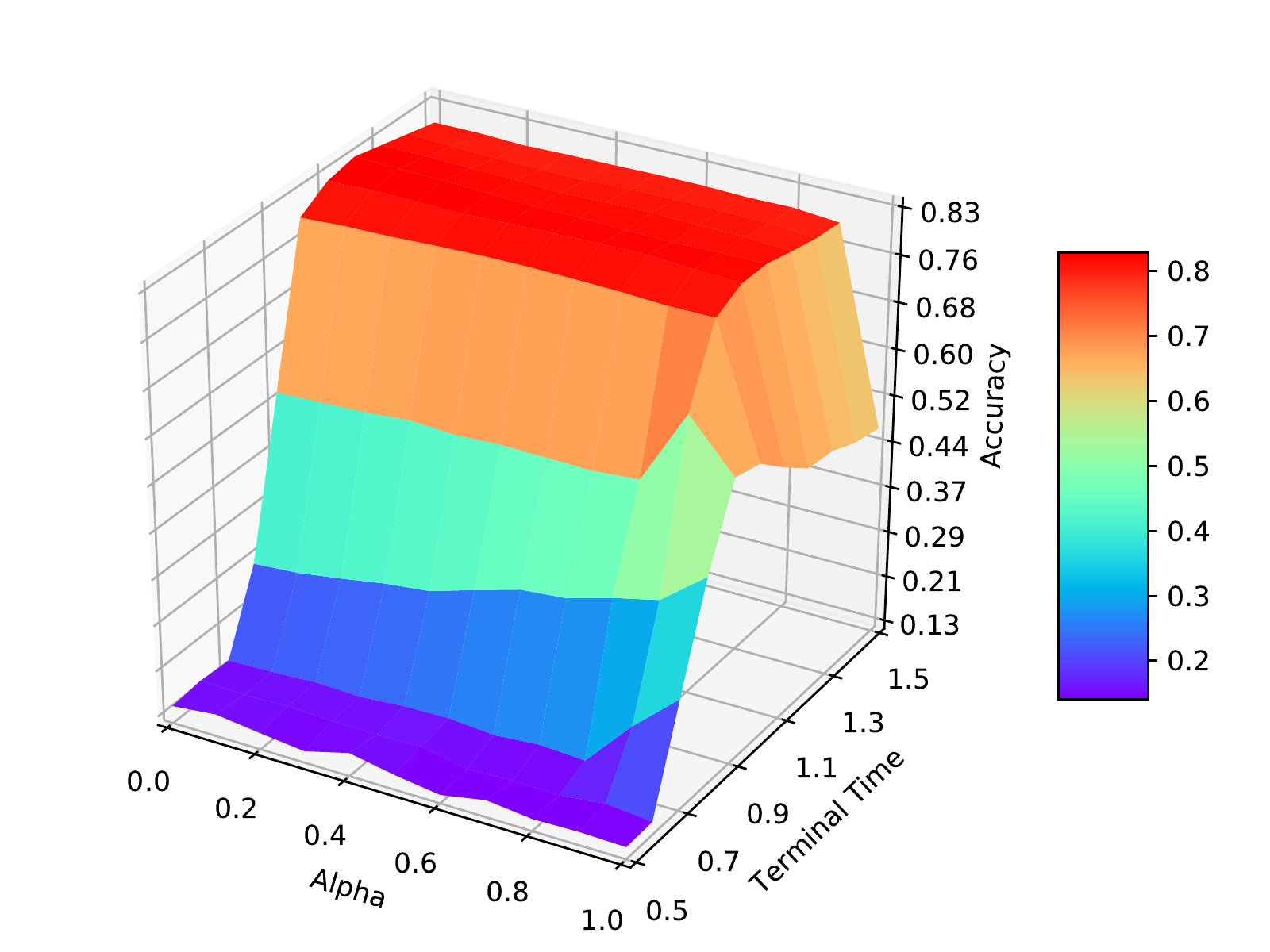}
\caption{  Mean classification accuracy of $100$ runs of our \model model over terminal time and $\alpha$ for the Cora dataset in 3D surface plot. }
\label{fig:appendix:cora3d}
\end{figure}

\begin{figure}[h]
\vspace{-0.5em}
\centering
\includegraphics[width=0.35\textwidth]{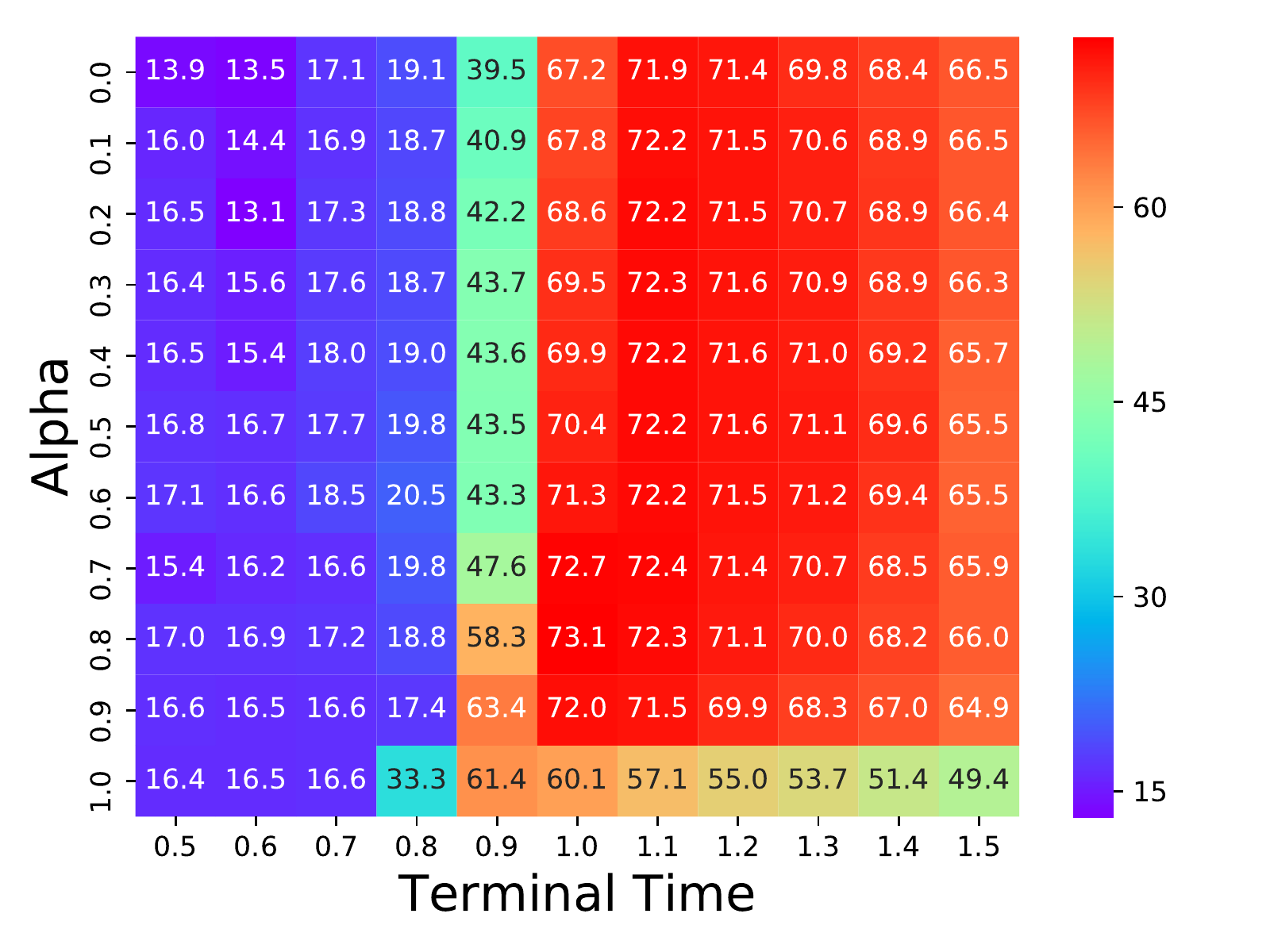}
\caption{  Mean classification accuracy of $100$ runs of our \model model over terminal time and $\alpha$ for the Citeseer dataset. }
\label{fig:appendix:citeseer}
\end{figure}

\begin{figure}[h]
\vspace{-0.5em}
\centering
\includegraphics[width=0.35\textwidth]{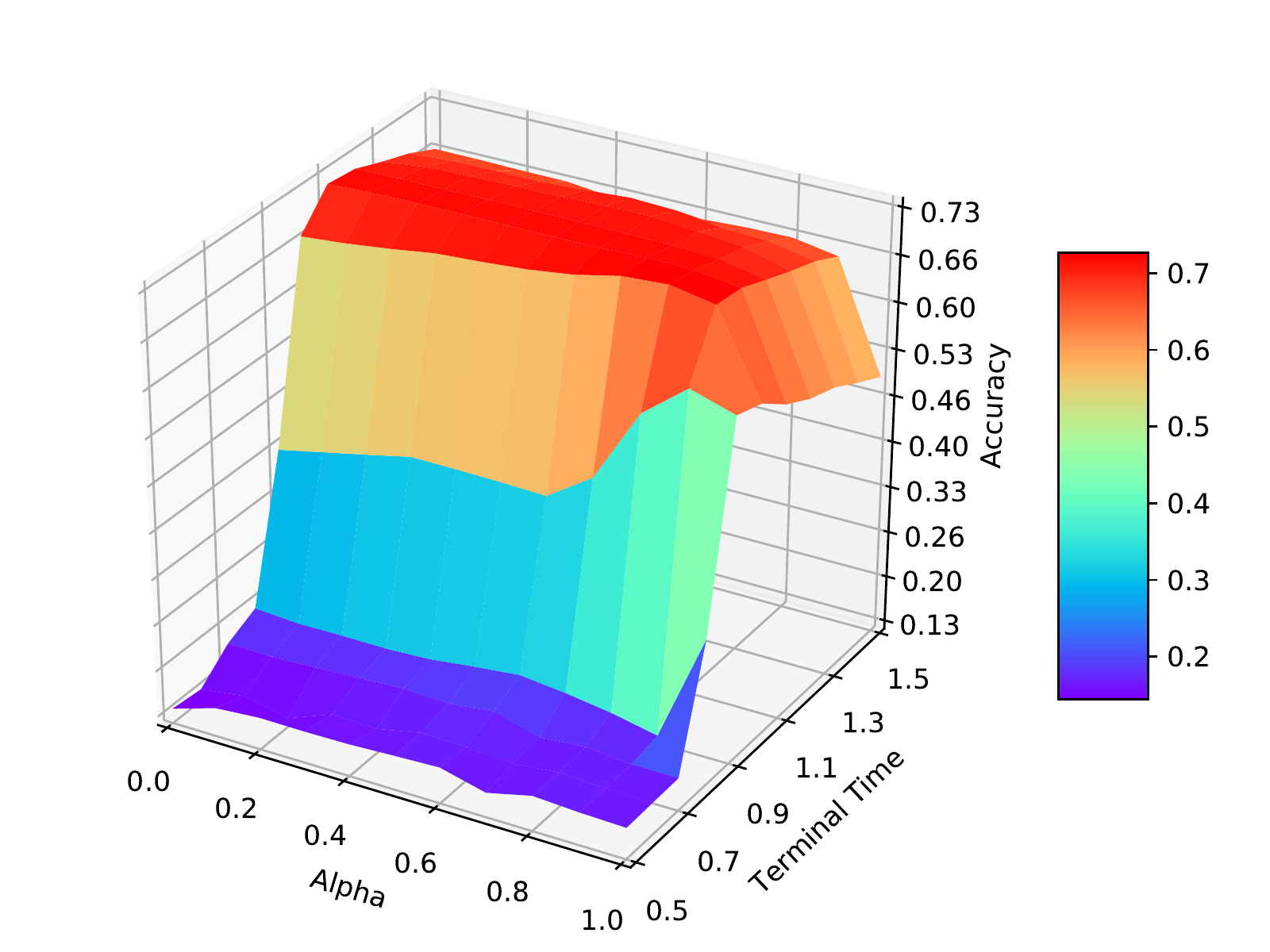}
\caption{  Mean classification accuracy of $100$ runs of our \model model over terminal time and $\alpha$ for the Citeseer dataset in 3D surface plot. }
\label{fig:appendix:citeseer3d}
\end{figure}

\begin{figure}[!ht]
\vspace{-0.5em}
\centering
\includegraphics[width=0.35\textwidth]{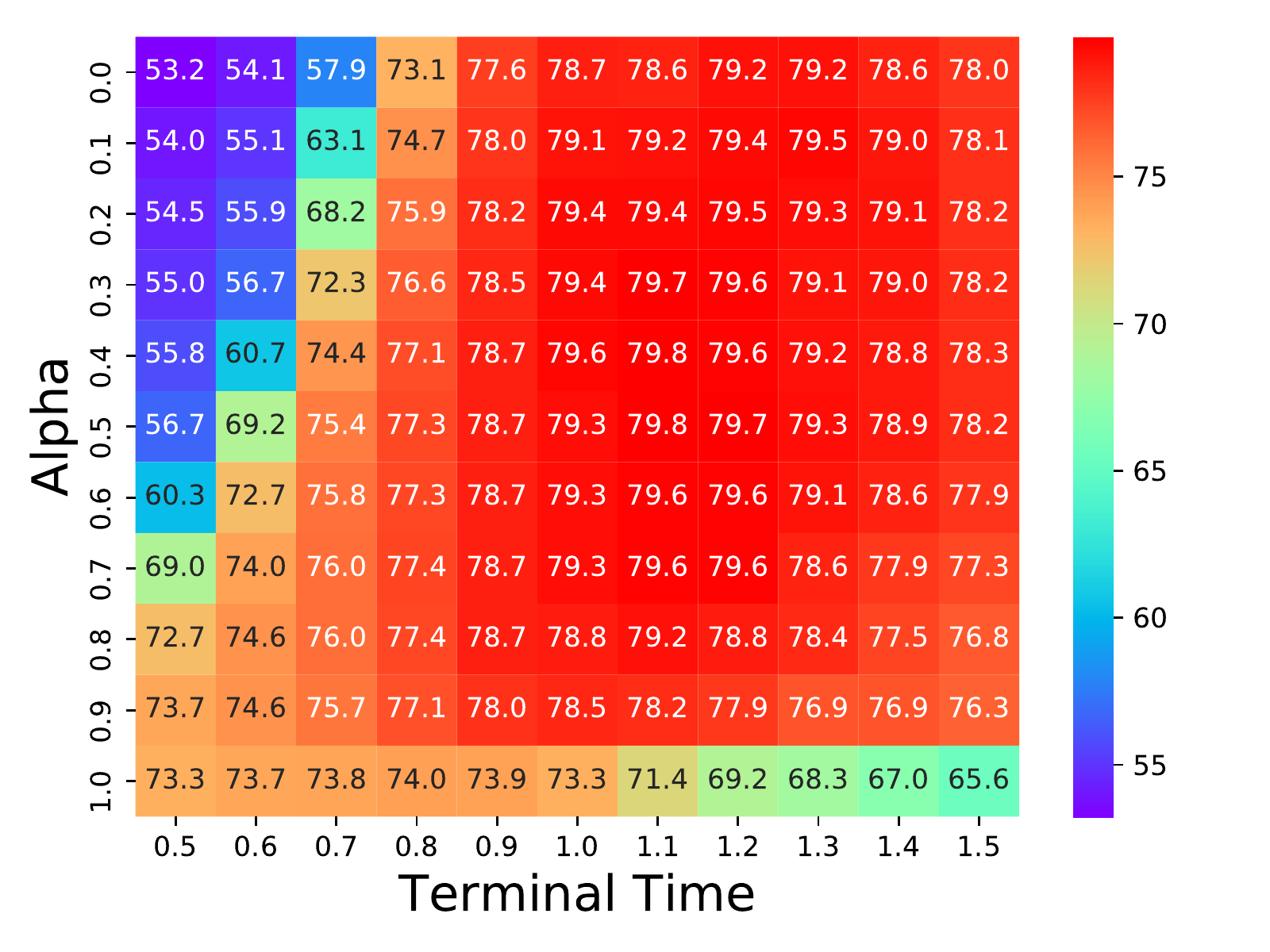}
\caption{  Mean classification accuracy of $100$ runs of our \model model over terminal time and $\alpha$ for the Pubmed dataset. }
\label{fig:appendix:pubmed}
\end{figure}

\begin{figure}[!ht]
\vspace{-0.5em}
\centering
\includegraphics[width=0.35\textwidth]{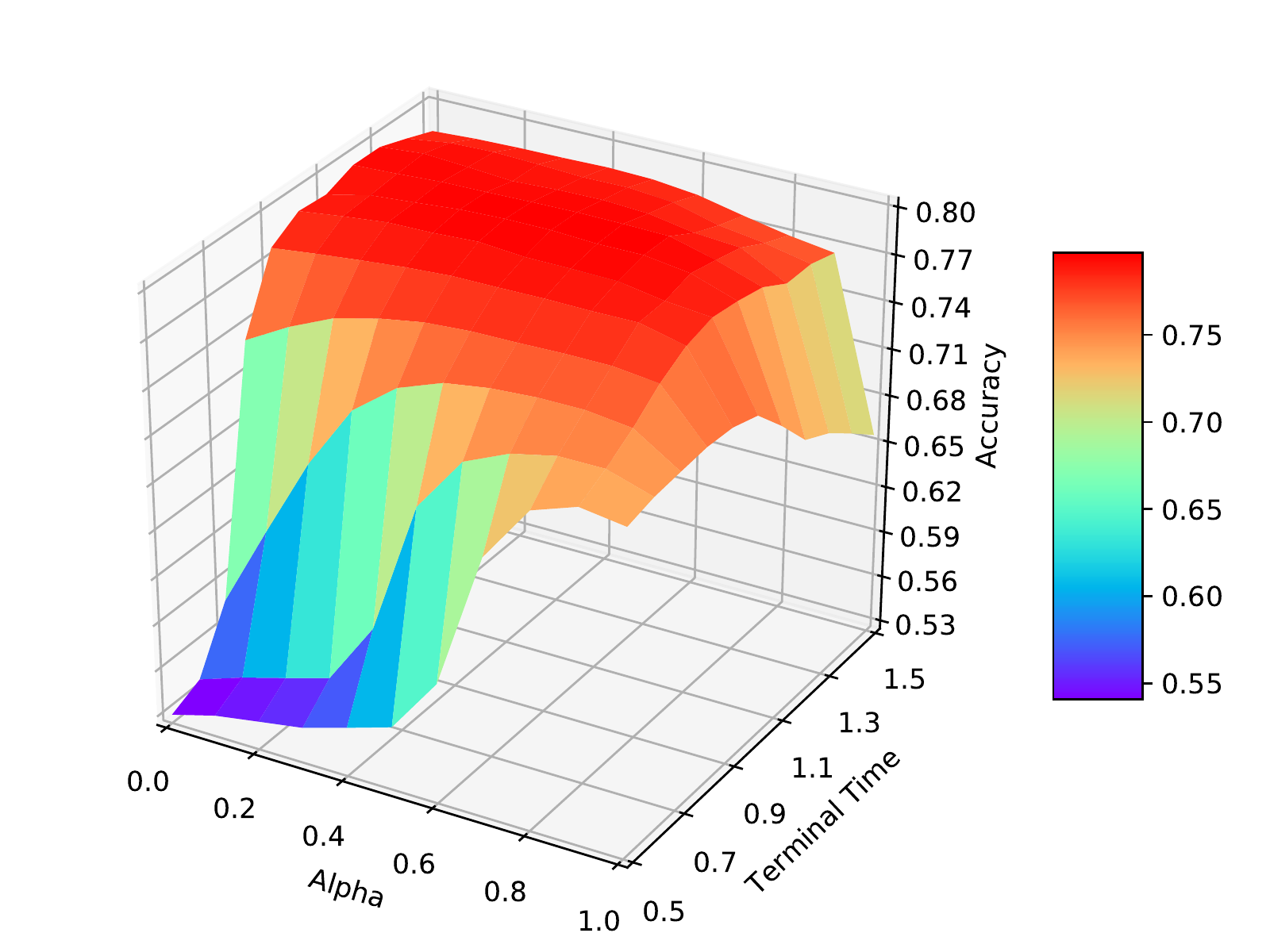}
\caption{ Mean classification accuracy of $100$ runs of our \model model over terminal time and $\alpha$ for the Pubmed dataset in 3D surface plot. }
\label{fig:appendix:pubmed3d}
\end{figure}

\end{document}